# Infall and disk processes – the message from meteorites

François L.H. Tissot[1,*], Christoph Burkhardt[2,*], Aleksandra Kuznetsova[3,4,5], Andreas Pack[6], Martin Schiller[7], Fridolin Spitzer[2], Elishevah M. M. E. Van Kooten[7], Teng Ee Yap[1]

[1]The Isotoparium, Division of Geological and Planetary Sciences, California Institute of Technology, Pasadena, CA 91125, USA (tissot@caltech.edu; tyap@caltech.edu).

[2]Max Planck Institute for Solar System Research, Department for Planetary Sciences, Justus-von-Liebig-Weg 3, 37077 Göttingen, Germany (burkhardtc@mps.mpg.de; spitzer@mps.mpg.de).

[3]American Museum of Natural History, Department of Astrophysics, 200 Central Park West, New York, New York, 10024 (akuznetsova@amnh.org)

[4]Center for Computational Astrophysics, Flatiron Institute, 162 5th Ave, New York, New York, 10010 (akuznetsova@flatironinstitute.org)

[5]University of Connecticut, Department of Physics, 196 Auditorium Rd, Storrs, Connecticut, 06269 (aleksandra.kuznetsova@uconn.edu)

[6]University of Göttingen, Geoscience Center, Department of Geochemistry and Isotope Geology, Goldschmidtstraße 1, 37077 Göttingen (apack@uni-goettingen.de).

[7]Center for Star and Planet Formation, Globe Institute, University of Copenhagen; Øster Voldgade 5-7, 1350 Copenhagen, Denmark (elishevah.vankooten@sund.ku.dk; schiller@sund.ku.dk)

*Correspondence to: tissot@caltech.edu and burkhardtc@mps.mpg.de

**Abstract (250 words)**

How do planetary systems, in general, and our own Solar System (SS), in particular, form? In conjunction, Astronomy and Isotope Cosmochemistry provide us with powerful tools to answer this age-old question. In this contribution, we review recent advances in our understanding of circumstellar disk evolution, including infall and disk processes, as explored through astrophysical models and nucleosynthetic isotope anomalies of SS materials.

Astronomically, filamentary structures and anisotropy are observed across the dynamic range of star formation and disk substructures are found to be ubiquitous, highlighting how star- and planet-forming environments are far more complex and dynamic than previously thought. Isotopically, two decades of investigation of nucleosynthetic anomalies in bulk meteorites and refractory inclusions have produced a rich dataset, revealing the existence of pervasive heterogeneity in the early SS, both at the large- (*i.e.*, NC-CC dichotomy) and fine-scale (*i.e.*, trends within the NC group). Using an updated data compilation, we review the systematics and emerging structures of these anomalies as a function of their nucleosynthetic origin. We present the two main families of models – *inheritance* vs *unmixing* – that have been proposed to explain the origin of the observed isotope heterogeneities, and discuss their respective implications for cloud infall and thermal processing in the disk. We also discuss how the extension of nucleosynthetic anomaly analyses to other chondritic components (Ameboid Olivine Aggregates, chondrules, matrix) has started to yield insights into transport, processing and mixing of dust in the disk. Limitations, open questions and key avenues for future work are presented in closing.

# Main Text

## 1. Introduction

From primitive mythological musings to more sophisticated Ptolemaic and Copernican theories, the origin of the Solar System (SS) and its planets has been amongst the most fundamental questions facing humankind. The discovery of exoplanetary systems has only made this question more relevant, offering a wealth of new samples to probe for comparison. Since the turn of the century, astronomical surveys have unveiled the remarkable diversity of planetary systems and broadly contextualized our SS within the galactic planetary census (Pinte et al. 2020; Thompson et al. 2018). From these observations emerges the view that our SS is statistically anomalous in its architecture (Rosenthal et al. 2022), suggesting the inadequacy of Astronomy alone in solving the essential mystery: how *did* our SS form? In this chapter, we review recent theoretical and experimental insights into the dynamics of SS formation, with a focus on those brought on by the study of meteorites.

To leading order, the astrophysics of planetary system formation is well understood. Indeed, despite limitations in the spatial resolution of astronomical observations, the plethora of systems available for study have shed light on the governing mechanisms that drive disk evolution. The onset of gravitational fragmentation within giant molecular clouds (GMC) initiates the collapse of dense cores, in which conservation of angular momentum leads to the formation of a disk that grows outwards from the central proto-star by viscous spreading. Within the forming disk, micron-sized dust particles grow to the mm- and cm-scale by hit-and-stick collisions (*e.g.,* Armitage 2020; Blum and Wurm 2008). Hydrodynamical processes, such as resonant drag instabilities (*e.g.,* Squire and Hopkins 2018; Youdin and Goodman 2005; see *Section 2*), then proceed to concentrate these particles into dust clumps, which will themselves collapse into ~100-km-scale planetesimals. Astronomical observations of star-forming regions and their protoplanetary systems provide snapshots of the structures that emerge from these phenomena, from GMC collapse and subsequent proto-cluster cloud formation (Friesen et al. 2016; Svoboda et al. 2019), to disk formation and evolution (Manara et al. 2022; Van Der Marel et al. 2021) (Fig. 1), and planet-disk interaction (*e.g.*, Keppler et al. 2018; Law et al. 2023).

Over the past decade, our understanding of the dynamics of star and planet formation has rapidly evolved due to (i) high spectral and spatial resolution studies in the (sub)-millimeter domain over a range of physical scales, as well as (ii) the increased dynamic range achieved by magneto-hydrodynamic (MHD) simulations of star formation from cluster to disk scales. The emerging astrophysical picture is that of a hierarchy of heterogeneous structures, present from star-forming filaments on molecular cloud scales (10's of parsecs) down to non-axisymmetric infalling flows called 'streamers' extending from core to planet-forming disk scales (< 0.1 pc - 10's of au) (see review in Pineda et al. 2022). In addition, interferometric surveys of planet-forming disks at various ages with multiple tracers with the Atacama Large Millimeter/submillimeter Array (ALMA) and the Very Large Array (VLA) have revealed that protoplanetary disks can be found with structured distributions of millimeter-sized dust concentrated in rings, gaps, spirals, or crescents, often differentially distributed to molecular gas (Andrews et al. 2018; Öberg et al. 2021; Ohashi et al. 2023; Tobin et al. 2020). These advances have raised exciting new questions, also relevant for the early SS, regarding the dynamical evolution of planet-forming disks, the transport and delivery of planetary ingredients, and the timing of planet formation.

In essence, our SS provides a test case for models of planetary system formation. While it is only one of a myriad of planetary systems in the Cosmos, the processes that shaped it can be mapped with exquisite detail through the study of meteorites. Indeed, meteorites act as witnesses to SS formation, preserving imprints of the physico-chemical processes at play in the earliest stages of its history. By inspecting the elemental and isotopic composition of meteorites and their components, one can probe the dynamic dialogue between the collapsing cloud and forming disk, constrain the physical, thermal, and chemical history of meteorite parent bodies, and trace the transport and mixing of materials between different disk reservoirs. The latter processes are most clearly revealed through the investigation of nucleosynthetic isotope anomalies. These anomalies reflect the heterogeneous distribution of products from stellar nucleosynthesis in the SS disk and, as elegantly stated by Clayton (1982), carry a *cosmic chemical memory*. Much like DNA – which made robust reconstructions of human migration patterns possible – nucleosynthetic anomalies hold the key to retracing the migration patterns of solids in the early SS, offering the opportunity to expose the primordial architecture of our SS and its evolution through time.

This contribution centers on recent developments concerning circumstellar disk evolution as explored through astrophysical models and nucleosynthetic isotope anomalies of SS materials. The former is discussed in *Section 2*, with emphasis on stellar birth environments, streamers, and the formation of disk sub-structures. In *Section 3*, we review the fundamentals and history of isotopic anomalies, and present their correlations in multi-element space. We then discuss how these anomalies can trace compositional changes in cloud infall in *Section 4a*, and in *Section 4b*, discuss how they may instead reflect transport and thermal processes in the disk. In *Section 5* we detail what can be learned about transport, processing and mixing of dust in the disk from isotopic anomalies in cometary materials and components of chondritic meteorites. Finally, we review the current limitations to our understanding of disk evolution and end with a discussion of open questions and an outlook in *Section 6*.

## 2. Astrophysical context
### *2.1 Structures of star-forming regions and protoplanetary disks*

Detailed observations of star-forming regions in the Solar neighborhood have revealed an array of potential templates on which to base our understanding of the birth environment of the SS (*e.g.*, Alves et al. 2020; Barclay et al. 2018), variable in evolutionary age (< 1 Myr in Ophiuchus to 5-10 Myr in Upper Sco), stellar density (10 - 1000 $M_\odot pc^{-3}$), or high energy radiation (UV to X-ray). These regions comprise a collection of young, protostellar systems, many of which host dust-rich planet-forming disks, inferred to have been born within a few Myr of each other and still embedded within their natal filamentary cloud. Studies of the dynamics and morphology of star-forming regions reveal that the natal gas distribution is quite heterogeneous: the larger scale filaments in which stars form were found to be composed of many different sub-filaments dubbed 'fibers', velocity coherent on relatively small scales (~ 0.1 - 0.3 pc) (Hacar et al. 2018) (Fig. 2). In numerical simulations that track star-formation from cluster to protostellar core scales, such structures are a natural consequence of the gravitational collapse of turbulent clouds. Recent studies find that assembly and accretion of protostellar material should be episodic and directional (Kuffmeier et al. 2017; Kuznetsova et al. 2019), reflecting the inherent heterogeneity of the star-forming environment. In these simulations (which have variable approaches to the treatment of feedback from massive stars), episodes of accretion onto nascent disks have been found to correspond to distinct spatial cloud reservoirs, suggesting a picture in which protostellar systems are supplied with fresh material from surrounding sub-filaments. Such filamentary flows of infalling cloud material, from 1000 AU - 10,0000 AU in length and extending down to disk scales, dubbed 'streamers' (Pineda et al. 2020), have been found around a number of sources in low-mass star-forming regions at various evolutionary stages. The dynamical timescales of observed streamers are relatively short (10's-100's of kyr) compared to the typical disk lifetime (3-10 Myr, Mamajek et al. 2004), suggesting that multiple episodes of infall during disk assembly and evolution are probable. For instance, the infalling streamer in Per-Emb 2 (Pineda et al. 2020) traced by $HC_3N$ emission, is kinematically distinct from a remnant envelope traced by $NH_3^+$ (Fig. 2 & 3). Notably, this type of anisotropic streaming infall in sources assumed to be at the earliest phases of disk evolution has been found to be coincident with annular structures, rings and gaps, in the millimeter dust emission profile of their planet-forming disks (Segura-Cox et al. 2020; Yen et al. 2019) (Fig. 2 & 3). Additionally, episodes of filamentary infall do not appear to be restricted to the early stages of disk formation, as several sources at later evolutionary ages (Class II) have been suggested to host remnant infalling envelopes or infalling streams (GM Aur: Huang et al. 2021, SU Aur: Akiyama et al. 2019).

While disk substructures have been identified in interferometric surveys targeting the thermal millimeter dust emission of nearby sources in younger embedded systems (Kido et al. 2023; Sai Insa Choi et al. 2023; Sheehan et al. 2020), they are particularly ubiquitous at later phases, when the disk is no longer as heavily enshrouded within an infalling envelope (Andrews et al. 2018). Dust concentration within substructures has been shown to be robustly produced when grains, drifting inward as they lose energy to gas drag, are arrested in their drift by a perturbation of the disk's pressure gradient (Pinilla et al. 2012). The presence of such perturbations, often referred to as dust traps, can have consequences for the long-term evolution of the disk structure and transport. Dust traps serve as both barriers between solid reservoirs, restricting the flow of solids, and as potential sites for efficient planet formation, locally enhancing the dust concentration and speeding up growth through coagulation and clumping mechanisms. Such perturbations, *i.e.*, pressure bumps, can be directly produced by the spiral wakes from the tidal forces of planets (Bae and Zhu 2018; Dong and Fung 2017; Zhu et al. 2015), at the edges of gaps opened by larger planets (Hammer et al. 2019; Zhu et al. 2012), or even due to hydrodynamic instabilities induced by infall itself (Kuznetsova et al. 2022). While there is no single agreed upon physical interpretation for the cause of observed disk substructures (see Bae et al. 2022; Morbidelli et al. 2024), their ubiquity among observed systems suggests that planet-forming disks are much more dynamic environments than previously thought.

### *2.2 Birth environment of the Solar System*

Taken together, observations of filamentary infall behavior and disk substructures have changed the picture of how and when material can be delivered to and transported within planet-forming disks. Of particular relevance to early SS formation and evolution is the viability of the delivery of compositionally heterogeneous material to the forming protosolar disk through infall (see *Section 4*). Streamers and other infall phenomena provide a compelling mechanism for how such material may be delivered, as well as for producing pressure perturbations potentially capable of keeping solid reservoirs separate during the evolution of the protosolar disk. Whether the SS was likely to have experienced such heterogeneous infall events in its early history and whether the composition of infalling material could adequately reproduce observed meteoritic trends may depend, in part, on the properties of the birth environment of the SS. We need to emphasize that the discovery and characterization of streamers is quite recent (*i.e.*, within the past 5 years), and most of the known sources at this point have been discovered serendipitously in existing ALMA data, not in any dedicated searches. As such, we keep in mind the aphorism "absence of evidence is not evidence of absence" when discussing the likelihood of streamers existing in specific types of environments.

The best-studied astrophysical systems are in the low-mass star-forming regions in the solar neighborhood. Because ALMA campaigns with adequate sensitivity and observational configurations to detect streamers have primarily focused on disks in such regions, most streamer detections come from low-mass star-forming regions. If the SS was born in such an environment, the available data already suggests that it would be likely to have experienced multiple episodes of infall during disk assembly and evolution. However, more than half of low-mass stars like our Sun are expected to form in high-mass star-forming regions (N>1000), and a number of works argue in favor of the Sun being born in such an environment, in close proximity to a massive star (Adams 2010; Desch and Miret-Roig 2024). External irradiation from a nearby massive star resulting in ~$10^4$ times the standard interstellar radiation field in the solar neighborhood today would result in a large degree of photoevaporation. If the protosolar disk evolved in a gas-poor environment cleared by external photoevaporation, akin to that of the proplyds (*i.e.*, photoionized protoplanetary disks) in Orion, accretion of fresh material through streamers or late-stage infall would be highly unlikely. Recent works, however, have used cosmochemical signatures such as the O and S isotopic ratios and the relative abundance of H, He, to Ar, Xe in Jupiter's atmosphere to constrain external irradiation levels from a nearby massive star to more moderate levels between 300 and 3,000 times the standard interstellar radiation field in the solar neighborhood today (Desch and Miret-Roig 2024; Monga and Desch 2014). The size of the SS is consistent with such far-UV fluxes, and the question is whether infall of compositionally heterogeneous material in such environment is likely. Given the bias of existing data towards low-mass star-forming regions, a proper discussion of likelihood is not possible. However, the recent detection of a streamer introducing compositionally variable material onto the Class I protostar Oph IRS 63 (Flores et al. 2023; Podio et al. 2024) shows that streamers do occur under moderate photoevaporation. Below, we examine whether enrichment of the early protosolar disk through infall is possible in scenarios in which the protosun is initially born in a more embedded environment or in proximity to massive stars.

Massive (OB) stars are core-collapse supernova progenitors, born in more densely packed massive star-forming regions such as Orion, rather than the nearby low-mass star forming regions inhabited by the most well-studied planet-forming disks. If the SS came from a massive star-forming region, it may have shared its birth environment with a SN progenitor responsible for injection of fresh material to the molecular cloud, including the Fe-peak elements and their nucleosynthetic anomalies. However, proximity to a supernova progenitor or a supernova explosion before and during SS formation is thought to result in hostile conditions for star formation and disk survival. If enrichment occurred early in the protosolar disk's lifetime and in close proximity to the progenitor (< 1 pc), as in the 'injection scenario', special conditions would be needed for the protosolar disk to survive photoevaporation from the strong FUV/EUV radiation emitted from the nearby massive star prior to explosion. Specifically, the protosolar system must have either (i) spent the majority of its time in the outskirts of a massive star-forming region (Adams 2010; Desch and Miret-Roig 2024) or (ii) been formed as a compact and massive enough system to withstand high amounts of external photoevaporation (Parker 2020). Considering direct injection through a more clumpy ejecta geometry loosens the constraints on the proximity of the SN progenitor (Ouellette et al. 2010) to a few parsecs. However, without additional shielding, at a few parsecs, elevated levels of FUV still place significant limits on the disk lifetime, and current models find that disk destruction occurs within 0.5-1 Myr for all but the most compact systems (Gárate et al. 2024).

Recent work argues that in an embedded environment, not only could the nascent molecular filament effectively shield the early SS from the effects of stellar feedback, but that streamers could themselves aid in the delivery of fresh materials (including short lived radionuclides) (Arzoumanian et al. 2023). Indeed, numerical simulations of star cluster formation that take into account shielding from high energy radiation by the surrounding gas and dust find much greater probabilities of disk survival for systems in massive star forming regions (Wilhelm et al. 2023). Recent observations of OMC-1, a region in close proximity (< 1 pc) to the massive star photoionizing the Trapezium region in Orion, find that the filamentary substructures in the protostellar environs (Hacar et al. 2018) are of such high density ($n_H > 10^7$ cm$^{-3}$), as to be largely unaffected by the nearby stellar feedback (Hacar et al. 2020).

Membership within a massive star-forming region is, however, not necessarily required for delivery of SN material. Indeed, not all massive stars remain within their natal environments, and an estimated 10-15 % of O stars are observed to be unbound from any star cluster (Maíz Apellániz et al. 2018). These 'runaways' are thought to be ejected from their host clusters through dynamical interactions. Estimates based on expulsion from binary systems due to the kick from the explosion of a more massive companion predict that O stars could travel 100s of parsecs from their birthplaces before their own SN event (Renzo et al. 2019). Furthermore, several works contend that there may exist some base level of enrichment in the natal gas due to self-pollution from massive (*e.g.*, Wolf-Rayet) stars and/or supernovae (Desch et al. 2022; Dwarkadas et al. 2017; Jacobsen 2005; Young 2014). If the SS formed in a scenario akin to that proposed for the nearby low-mass star-forming regions of the Local Bubble – in which star formation was triggered by the expansion of SN ejecta into the interstellar medium (Zucker et al. 2022) – enrichment and formation of the SS birth cluster may very well go hand in hand (Gounelle and Meynet 2012). Compression from the pressure-driven expansion of SN ejecta can trigger star formation in marginally stable clouds, but enriched ejecta from the SN explosion must cool before it can be incorporated into star-forming material. Numerical simulations tracking SN material within turbulent star-forming molecular clouds find that hot ejecta do not mix efficiently with the existing cool molecular material. Enriched material is thus primarily incorporated after it has had time to cool enough to be accreted by protostellar systems after their initial phase of disk formation (t > 100 kyr) (Kuffmeier et al. 2016). Recent observations of streamer sources in the NGC 1333 star-forming region connect the smaller-scale inflow onto protostellar sources to large-scale flows originating outside of the host filament (Valdivia-Mena et al. 2024), supporting a picture in which early phase disks can be expected to experience later-stage accretion of compositionally distinct material.

These recent insights suggest that it is important to consider that the time at which compositionally distinct material is accreted onto the protosolar disk does not necessarily correspond to the timing of an enrichment event like a SN explosion. The composition of accreted material can also change over time when protostars move through and accrete from spatial reservoirs of varying composition. Numerical simulations find that protostars accrete material from well outside the bounds of the initial cores, in some cases traveling on the order of 0.3 - 1.0 pc from their initial formation location within < 1 Myr (Kuznetsova et al. 2015). Observed SN remnants show spatial compositional heterogeneity (*e.g.*, Holland-Ashford et al. 2020; Winkler et al. 2014), suggesting that anisotropies in the ejecta are imprinted on the SN remnants structure. Recent 3D simulations of SN explosions have even been able to capture the development of these anisotropies across scales (Mandal et al. 2023), even accounting for variable yields of Ti/Fe in ejecta compared to spherically symmetric models (Sieverding et al. 2023).

Two important constraints when considering the birth environment of the protosolar disk are (i) the source(s) and (ii) the timing of the enrichment of the natal gas in short-lived radionuclides

(SLR) and other *r-/s-* process elements through stellar feedback: both through winds driven by massive stars and in the ejecta of core-collapse supernovae. Short-lived radionuclides such as $^{26}$Al ($t_{1/2}$ = 0.7 Myr) inform on the relative timing of enrichment, as their very presence in the early SS requires that they were produced and injected into the parental cloud shortly before the formation of the SS (*e.g.*, Dauphas and Chaussidon 2011; Tissot et al. 2016; Young 2014). If present, isotopic heterogeneity for SLR would additionally require that the enrichment itself is spatially or temporally heterogeneous. Although there is some debate about the (small-scale) homogeneity of $^{26}$Al (*e.g.*, Desch et al. 2023c; Krestianinov et al. 2023) and $^{92}$Nb (Hibiya et al. 2023), the remarkable concordance of various SLR chronometers among different planetary bodies points to a homogeneous distribution of SLRs in the ESS to within a level of a few per cent (*e.g.*, Dauphas and Chaussidon 2011; Davis and McKeegan 2014; Desch et al. 2023c; Desch et al. 2023b; Nyquist et al. 2009). By contrast, isotopic anomalies of non-radiogenic, nucleosynthetic origin are observed among different ESS materials in the parts per ten thousand to parts per million range (ε to μ levels). Even though these isotopic variations are tiny, they provide essential information about the formation and evolution of the solar system.

In summary, star formation is a multi-scale and multi-physics problem of immense dynamic range. Our theoretical understanding has evolved in order to account for how the large-scale environment affects the formation and evolution of planetary systems, resulting in a much more dynamic and complex picture than previously thought (Fig. 2). While the birth environment of the SS remains uncertain, constraints on the timing and relative amounts of compositional variation traced by meteoritic evidence are available. These could help constrain the likelihood of various enrichment scenarios and/or environmental conditions relevant to the early SS, and establish its context within the greater Galactic environment.

## 3. Isotope anomalies in meteorites
### *3.1 Basics and history*
#### *(i) What are isotope anomalies?*

As samples of planetary bodies forming in the solar accretion disk, meteorites are key for constraining the origin, makeup, and physico-chemical evolution of matter in the Sun's birth environment. Investigating isotope variations in meteorites and their components is particularly useful in this respect. For an insightful discussion, it is important to realize that isotopic variations are classified into two broad categories: *mass-dependent* and *mass-independent*. As the name indicates, *mass-dependent* isotope effects are those in which the degree of isotopic fractionation scales proportionally with isotopic mass differences. These effects arise during equilibrium processes in (iso-)chemical exchange reactions due to differences in the zero-point energy of isotopically substituted molecules, as well as during kinetic (*i.e.,* unidirectional) processes as a result of the faster reaction and/or diffusion of the lighter isotopes relative to the heavier ones. Mass-dependent isotope effects are generally large (‰ level or more) and can provide insights into the conditions of phase equilibria, or processes like evaporation, condensation, and diffusion. Extensive description of mass-dependent effects, including their quantum mechanical origin, can be found in the literature (*e.g.*, Albarede et al. 2004; Bigeleisen and Goeppert-Mayer 1947; Dauphas and Schauble 2016; Ibañez-Mejia and Tissot 2021; Marechal et al. 1999; Urey 1947; Young et al. 2002).

In contrast, *mass-independent* effects are those that *do not* scale proportionally with the mass difference between isotopes but relate to other nuclide properties like radioactivity, neutron capture cross section, or radius. These effects present themselves as deviations from mass-dependent behavior

(Fig. 4). In order to isolate them, the measured data need to be corrected for natural and analytical mass-dependent effects. Such corrections can only be done for elements with at least three stable isotopes, which unfortunately means that for elements like H, C and N (three key volatile elements), differentiating between mixing and material processing is extremely arduous. For elements with 3 isotopes or more, correction of mass-dependent effects is typically performed using an internal normalization approach, whereby the degree of mass-dependent fractionation (often denoted β) experienced by the sample is determined by (i) fixing the "normalizing ratio" to a reference value (most often the terrestrial composition), and (ii) assuming a particular mass-dependent law (typically, the exponential law) (*e.g.,* Marechal et al. 1999; Russell et al. 1978). The β value thus obtained is then used to remove the effect of mass-dependent fractionation from the isotope ratios of interest. Residual departures from mass-dependency after internal normalization are mass-independent effects.

Not all mass-independent effects are considered "isotope anomalies", a term reserved here for isotopic variations that cannot be explained by radiogenic ingrowth, cosmo- and nucleogenic production, or Nuclear Field Shift effects (Fig. 4). These isotope anomalies are typically small, and most often reported in epsilon (ε) and mu (μ) notations, which are, respectively, part-per-ten-thousand and part-per-million (ppm) deviations from the terrestrial composition. At high precision (*i.e.,* ppm level), spurious effects can arise from a host of analytical effects, requiring that great care be taken to measure samples and standards under identical conditions. Similarly, for samples that have experienced large degrees of mass-dependent fractionation, correction with an inappropriate mass-fractionation law can result in apparent anomalies, which can mislead data interpretation (*e.g.*, Budde et al. 2023; Davis et al. 2015; Olsen et al. 2013; Zhang et al. 2014). In this contribution, we will focus on two types of isotopic anomalies: (i) those affecting oxygen isotopes, the origin of which remains debated but is most often understood as a result of isotope-selective photodissociation of CO in the solar nebula (Clayton 2002), and (ii) those of nucleosynthetic origin, which reflect the heterogeneous distribution of isotopically anomalous presolar grains in the solar nebula (*e.g.*, Dauphas and Schauble 2016).

*(ii) Initial discoveries to present golden age*

Historically, it was assumed that all SS materials had the same initial isotopic composition as a result of homogenization in a hot solar nebula (Cameron 1962). A string of discoveries spanning the last five decades have since falsified this notion, revealing the existence of isotopic heterogeneity in SS materials from the micron- to the planetary-scale. The history of this paradigm shift regarding the chemistry of the SS is well summarized in Zinner (2014) and Dauphas and Schauble (2016). In brief, the first evidence for isotopic heterogeneity of the solar nebula was discovered in the noble gases Xe (Reynolds & Turner, 1964) and Ne (Black 1972; Black and Pepin 1969) extracted from chondrites (chondrites are meteorites from asteroidal parent bodies that did not experience sufficiently severe thermal metamorphism to melt, and thus can preserve original disk material). Subsequent work involving the sequential digestion of chondrites eventually led to the identification of nanodiamonds (Lewis et al. 1987), SiC grains (Bernatowicz et al. 1987; Tang and Anders 1988), and graphite grains (Amari et al. 1990) as the presolar carriers of these isotope anomalies. This revealed that grains forming in the outflows of dying stars survived the formation of the SS, and provided a benchmark to test models of stellar nucleosynthesis on natural samples in the lab. Since then, the use of advanced microanalytical techniques (*e.g.,* Nanoscale Secondary Ion Mass Spectrometry, or Nano-SIMS; Resonance Ionization Mass Spectrometry, or RIMS), has led to the identification of a variety of additional presolar phases, including oxides and silicates, and yielded isotope anomaly data for an ever increasing number of elements and stellar sources (*e.g.*, Stephan et al. 2024).

The discovery of isotope anomalies in major rock-forming elements came only shortly after the finding of variations in the noble gases and began with O (Clayton et al. 1973), before rapidly expanding to Mg, Ca, Ti, Cr, and Ba (*e.g.*, Clayton et al. 1988 and references therein). These anomalies were found in Ca-Al-rich inclusions (CAIs; sub-mm to cm-sized highly refractory ceramic-like aggregates present in chondrites; their mineralogy is consistent with the one predicted for the first solids condensing from a cooling gas of solar composition, Grossman 1975; Grossman 1972) and are smaller than the ones in the presolar grains. This is an effect of averaging and dilution, which also makes it challenging to link the anomalies in CAIs to specific carriers and stellar sources. Indeed, not all presolar carriers are refractory phases that can survive chemical digestion with strong acids like SiC, nanodiamonds, or graphite, and painstaking workflows involving physical and chemical separation protocols are needed to identify and study these less-refractory carriers. A notable success in the application of such an approach has been the identification of nano-spinels as the carrier of $^{54}$Cr anomalies (*e.g.*, Dauphas et al. 2010; Qin et al. 2010).

With advancements in mass spectrometry, more subtle anomalies became resolvable. In particular, the development of Multi-Collector Inductively Coupled Plasma Mass Spectrometry (MC-ICPMS) instruments has revealed that isotopic anomalies of refractory elements are not confined to presolar grains and meteorite inclusions, but also present in bulk meteorites and samples of planetary-sized bodies (*e.g.*, the Moon, Mars, Earth). Dauphas et al. (2002a) first showed that nucleosynthetic Mo anomalies persisted in magmatic iron meteorites, which are thought to sample the metallic cores of differentiated planetary bodies (10s to 100s of km in diameter). By picking up on earlier work (Niemeyer 1988a; Rotaru et al. 1992), widespread heterogeneity was then also described in the Cr, Ti, and Ni isotope composition of bulk chondrites and achondrites (*e.g.*, Leya et al. 2008; Regelous et al. 2008; Trinquier et al. 2009; Trinquier et al. 2007). These results triggered a zealous search for nucleosynthetic anomalies in numerous refractory elements (see Dauphas and Schauble 2016), and have ushered in a golden age for the field (Fig. 5). Today, nucleosynthetic isotope anomalies have been documented in elements spanning a wide range of geo- and cosmochemical behaviors: from lithophile (*e.g.*, Ca, Ti, Cr, Ba) to siderophile (*e.g.*, Fe, Mo, Ni, W), and ranging in volatility from highly refractory (*e.g.*, Ca, Ti, Sr, Zr) to moderately volatile (K, Zn) (see Table 1). While the carrier phases of almost all these anomalies remain enigmatic, their heterogeneous distribution in SS materials can be leveraged to understand the evolution of the early SS.

*(iii) Isotope anomalies as tracers of SS formation*

Forged under the extreme temperatures, pressures, and neutron fluxes of stellar environments, nucleosynthetic signatures are essentially immune to modification by physico-chemical processes taking place in interstellar media and circumstellar disks. As such, the presolar grains present in the SS parent molecular cloud are *de facto* inert and robust tracers of its collapse and infall onto the nascent protoplanetary disk, as well as the transport and mixing of materials therein. The end product of these complex processes is the current distribution of isotope anomalies in SS materials, which reflect different contributions of nuclides produced in different stages of stellar evolution and astrophysical settings (*e.g.*, Burbidge et al. 1957; Cameron 1957; Clayton 1983). Most relevant to the subject of nucleosynthetic anomalies discussed here are the so-called Fe-peak elements (*e.g.*, Ti, Cr, Fe, Ni, Zn) and the elements with atomic masses beyond ~80 (*e.g.*, Sr, Zr, Mo, Ru, Nd). Iron-peak elements are produced in neutron-rich environments either at or near nuclear statistical equilibrium (NSE) in massive stars ($\geq 8$ M$_\odot$) prior and during core-collapse (Type-II) supernovae, as well as in rare Type Ia supernovae, which are thought to result from binary stellar evolution (Heger et al. 2014). For some Fe-peak elements, anomalies are attributed to excesses in neutron-rich isotopes (*e.g.,* $^{54}$Cr,

$^{50}$Ti, $^{48}$Ca) (Dauphas et al. 2014; Trinquier et al. 2009; Trinquier et al. 2008). For others, anomalies reflect excesses in the neutron-poor isotopes ($^{58}$Ni, $^{54}$Fe) (Hopp et al. 2022a; Hopp et al. 2022b; Steele et al. 2011). All elements beyond the Fe-peak are produced via either the *s*- , *r*- , and/or *p*-process, which respectively stand for 'slow neutron capture', 'rapid neutron capture', and 'proton-capture' or 'photodisintegration' (Burbidge et al. 1957; Cameron 1957). The *s*-process occurs in intermediate-mass stars (0.5–8 M$_\odot$) during the asymptotic giant branch (AGB) phase of their evolution (Gallino et al. 1998). For instance, variation in *s*-process contributions to SS materials is thought to be reflected in isotope anomalies in the *s*-peak elements Zr, Mo, and Ru (Akram et al. 2013; Burkhardt et al. 2011; Chen et al. 2010). The *r*-process occurs in environments characterized by extremely high neutron fluxes, such as supernovae and neutron star mergers. The *p*-process (more often named *γ*-process) is less well understood, and a multiplicity of sites have been proposed, including the O/Ne-shell of exploding massive stars, white dwarf explosions, or thermonuclear burning on the surface of neutron stars (Rauscher et al. 2013).

By quantifying the minute differences in isotope anomalies between meteorites and planetary bodies at high precision, isotope cosmochemists have been able to place powerful constraints on early SS dynamics and planetary formation. Most notably, these anomalies have revealed that for virtually all elements, bulk planetary materials can be classified into two isotopically distinct supergroups formed early in, and preserved across, SS history. These two supergroups, termed non-carbonaceous (NC) and carbonaceous (CC), are thought to broadly correspond to inner and outer SS materials, respectively, with their separation conventionally attributed to either (i) the early formation of proto-Jupiter close to the water-ice sublimation front (Burkhardt et al. 2019; Kleine et al. 2020; Kruijer et al. 2017; Warren 2011), or (ii) preferential planetesimal formation at the silicate and/or water sublimation fronts, with Jupiter playing a secondary role (Lichtenberg et al. 2021; Morbidelli et al. 2022). The existence of such isotopic heterogeneity has provided unprecedented opportunities to probe the history of the SS and its materials. For instance, leveraging differences in the siderophilic affinities of elements, nucleosynthetic anomalies have been shown to track the provenance of planetary building blocks across accretion of the terrestrial planets (*e.g.*, Dauphas 2017). Recently, the more volatile behavior of K and Zn has also been used to investigate the origin of terrestrial and Martian volatiles (Kleine et al. 2023; Martins et al. 2023; Nie et al. 2023; Savage et al. 2022; Steller et al. 2022).

Key to the interpretation of the isotopic heterogeneity observed in SS materials is the question of its origin. Were the presolar carriers themselves heterogeneously distributed in the parent molecular cloud, such that an isotopically heterogeneous disk is merely a reflection of compositional changes in infalling material (Burkhardt et al. 2019; Nanne et al. 2019; Yap and Tissot 2023) (see *Section 4.1*)? Or on the contrary, were the carriers homogeneously distributed in the cloud, with the observed heterogeneity arising from physical (*e.g.,* dust grain size-sorting) and/or thermal processing during disk evolution (*e.g.,* Burkhardt et al. 2012; Dauphas et al. 2010; Trinquier et al. 2009) (see *Section 4.2*)? Similarly, in the search for planetary building blocks, recent work has hinted at the existence of a "missing" building block for the Earth that is not sampled by bulk meteorites (Burkhardt et al. 2021). Where is this material and why has it eluded detection? These questions have sparked lively debate in the field. As an illustrative example, in the past two years there have been studies challenging the existence of an NC-CC dichotomy between SS materials (Onyett et al. 2023), supporting it (*e.g.,* Yap and Tissot 2023), and positing the existence of an NC-CC-CI *trichotomy* instead (Dauphas et al. 2024; Hopp et al. 2022a). In this contribution, we aim to present the available data and the opposing views they have kindled, focusing on the overarching question: *What do isotope anomalies tell us about infall and/or disk processes?*

### *3.2 Oxygen isotope anomalies*

A key rock forming element, oxygen has three stable isotopes ($^{16}O$, $^{17}O$, $^{18}O$). These display some of the largest, non-radiogenic, anomalies observed in SS bodies and have been historically critical in the identification of genetic links between extraterrestrial materials (see also review by Ireland et al. 2020, and references therein). The processes at the origin of O isotope anomalies, however, are distinct from those envisioned for other elements, *e.g.*, Cr or Ti (see below). Furthermore, apart from mass-independent fractionation, mass-dependent fractionation and exchange between reservoirs (*e.g.*, gas and dust) play a role in setting the O isotopic composition of meteoritic materials. Here we use the classical $\delta^{17}O$ and $\delta^{18}O$ notations along with the $\Delta'^{17}O$ notation as introduced by Miller (2002). The $\Delta'^{17}O$, which is defined as:

$$\Delta'^{17}O = ln\left(1 + \frac{\delta^{17}O_{SMOW}}{1000}\right) - 0.528\, ln\left(1 + \frac{\delta^{18}O_{SMOW}}{1000}\right) \quad (1)$$

is synonymous with the "$^{17}O$-excess" parameter used by the water community.

*(i) Discovery and proposed origins/mechanisms*

Clayton et al. (1973) first discovered O isotope anomalies in refractory CAIs in carbonaceous chondrites and observed that the data fell on a slope ~1 line in $\delta^{17}O$ vs. $\delta^{18}O$ space. Back then, the O isotope composition of silicates was analyzed by fluorinating rocks and conversion of the liberated $O_2$ to $CO_2$ for mass spectrometric analysis. Clayton et al. (1973) observed a correlation between the apparent $\delta^{13}C$ and the $\delta^{18}O$ of the analyzed $CO_2$ and correctly attributed the shift in $\delta^{13}C$ to mass-independent fractionation in $\delta^{17}O$. Later, direct analysis of the liberated $O_2$ became common and allowed measurements of $\Delta'^{17}O$ with much higher precision.

Different processes have been presented to explain O isotope anomalies. Originally, Clayton et al. (1973) proposed that the bulk SS has a composition similar to that of the Earth ($\delta^{18}O$ = +5.5 ‰, $\Delta'^{17}O$ = –0.05 ‰, Pack et al. 2016) and suggested a nucleosynthetic origin of O isotope anomalies in $^{16}O$-rich CAIs (see also Clayton et al. 1977). The large, differentiated inner SS bodies from which we have samples (Earth, Moon, Mars, Vesta), all have O isotope compositions defining a tight cluster around $-0.3 \leq \Delta'^{17}O \leq +0.3$ ‰. Without any measurements and following Occam's razor, one would conclude that the bulk SS is at $\delta^{18}O \approx$ +5 ‰ and $\Delta'^{17}O \approx$ 0 ‰. The question of the bulk O isotope composition of the SS has since been addressed by measuring solar wind particles implanted in lunar regolith, however this returned ambiguous estimates of $^{16}O$-rich (Hashizume and Chaussidon 2005) and $^{16}O$-poor compositions (Ireland et al. 2006). The dispute was resolved when the results from the GENESIS mission became available (McKeegan et al. 2011) and, independently by spectroscopic measurements of the solar photosphere (Lyons et al. 2018). The data from GENESIS were interpreted to show mass-dependent fractionation and projection of the data on the slope-1 line defined by CAIs yielded $\delta^{17}O = \delta^{18}O = -58$ ‰ ($\Delta'^{17}O = -28$ ‰). The direct measurements of the solar photosphere gave $\delta^{18}O = -50 \pm 11$ ‰ (Lyons et al. 2018).

The spectroscopic data support the correction of the GENESIS data and confirm that the SS is $^{16}O$-rich relative to asteroids and planets. These observations contradict the interpretation of CAIs having incorporated anomalous $^{16}O$-rich material and that the bulk SS is at $\Delta'^{17}O \approx$ 0 ‰. If the CAIs had incorporated material with nucleosynthetic anomalies for O, one would also expect large, correlated isotope anomalies with other elements. In one early study, Clayton and Mayeda (1977) reported on two CAIs with anomalous oxygen, for which the anomaly correlated with the anomaly in

Mg isotopes (Wasserburg et al. 1977), which has been traced back to the presence of short-lived $^{26}$Al (Lee et al. 1977). However, correlated anomalies for a larger set of samples and elements have not been observed (Clayton 1993; Clayton 1978) and the nucleosynthetic origin of the O isotope anomalies in bulk meteorites has since largely been dismissed.

Thiemens and Heidenreich (1983) and Heidenreich and Thiemens (1983) experimentally observed a novel chemical process that leads to O isotope fractionation similar to that observed in meteorite components. Their experiments on ozone formation led to the formation of $O_3$ with large positive $\Delta'^{17}O$ and $O_2$ with negative $\Delta'^{17}O$. Educt $O_2$ and reaction product $O_3$ fall on a slope-1 line in a $\ln(\delta^{17}O + 1)$ vs. $\ln(\delta^{18}O + 1)$ space, just like refractory inclusions in carbonaceous chondrites do. The mass-independent fractionation is due to symmetry effects in the O-O-O molecule and differences in reaction rates. An $^{16}O_3$ reaction product tends to decompose faster from the excited state back to $O_2$ whereas asymmetric $O_3$ with substituted $^{17}O$ or $^{18}O$ has a longer half-life making it more likely to react to stable $O_3$. The authors suggest that the formation of symmetric molecules containing oxygen (*e.g.*, $SiO_2$, $TiO_2$) could be responsible for the fractionation along a slope-1 line observed in meteorites (see also volume η [Greek "eta"] effect, Marcus 2004).

A different type of chemical reaction that produced isotope anomalies was suggested by Marcus (2004). He proposed that the equilibration between adsorbed monoxides ($XO_{ads}$) and oxygen atoms ($O_{ads}$) and excited dioxides ($XO_{2,ads}^*$, Eq. 2) can lead to mass-independent effects, *e.g.*, during condensation of dust in the early solar nebula:

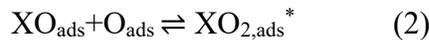
$$XO_{ads}+O_{ads} \rightleftharpoons XO_{2,ads}^* \qquad (2)$$

The element "X" in Eq. 2 could be Si, Al, Ti, or other metals. Such surface η effect yields $XO_{2,ads}^*$ that is mass-independently enriched in $^{17}O$ and $^{18}O$ and $XO_{ads}$ and $O_{ads}$ that are depleted in $^{17}O$ and $^{18}O$ (Eq. 2). The excited $XO_{2,ads}^*$ evaporates and reacts with $H_2$ to form $^{16}O$-poor $H_2O$, whereas the adsorbed XO and O react with adsorbed metal atoms (*e.g.*, Ca) or monoxides into $^{16}O$-rich oxides and silicates of the CAIs. Chakraborty et al. (2013) experimentally demonstrated that the oxidation of gaseous SiO, which is the major Si-bearing component in a gas of solar composition at high temperatures, by OH leads to formation of solid $SiO_2$ (Eq. 3), which is selectively depleted in $^{16}O$. This could explain the $^{16}O$-poor composition of silicates (*e.g.*, chondrules) relative to the $^{16}O$-rich composition of the solar nebula without a nucleosynthetic $^{16}O$ source (and CO self-shielding).

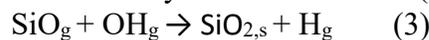
$$SiO_g + OH_g \rightarrow SiO_{2,s} + H_g \qquad (3)$$

The reaction shown in Eq. 3 may go through a vibrationally excited transition state (SiOOH*). As in the case of $O_3$ formation, the deactivation of the intermediate state would then be the place of mass-independent fractionation (Chakraborty et al. 2013).

A different perspective on the origin of the mass-independent fractionation in meteorites and their components has been published by Clayton (2002). He suggested that predissociation and self-shielding of CO leads to formation of $^{16}O$-poor $H_2O$. Isotope separation by self-shielding of CO had been observed in molecular clouds (Bally and Langer 1982). Those photons that predissociate $C^{16}O$ are quickly consumed and, at greater penetration depths, only $C^{17}O$ and $C^{18}O$ are pre-dissociated. The O radical reacts with ambient $H_2$ to $H_2O$ (Eq. 4). This $H_2O$ is then selectively, and by the same factor, enriched in $^{17}O$ and $^{18}O$.

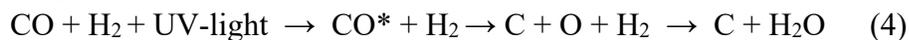
$$CO + H_2 + UV\text{-light} \rightarrow CO^* + H_2 \rightarrow C + O + H_2 \rightarrow C + H_2O \qquad (4)$$

Silicates and oxides that react with such $H_2O$ would then fall along a slope-1 line in the three-oxygen isotope space. The different O isotope fractionation regimes are illustrated in Fig. 6 (modified after Thiemens and Lin 2021). Originally, Clayton (2002) suggested that the source of the UV radiation is the young Sun and that self-shielding occurred in the inner SS. High ambient temperatures, however, would lead to quick back-equilibration and disappearance of the anomaly. Lyons and Young (2005) modeled self-shielding with an external UV source (O or B star within 1 pc) and location of self-shielding at the disk surface. The environment in the outer, cool parts of the disk may resemble the conditions in cold molecular clouds. Yurimoto and Kuramoto (2004) placed the origin of the mass-independent O isotope fractionation in the cold molecular cloud from which the SS formed. In such an environment, dissociation of CO is followed by surface-mediated reaction of O with $H_2$ on the cold surfaces of grains. The ice is $^{17}O$ and $^{18}O$ rich, whereas the CO is $^{16}O$ rich. These studies (Clayton 2002; Lyons and Young 2005; Yurimoto and Kuramoto 2004) have in common that they predict that the SS (*i.e.*, the Sun) has a $^{16}O$-rich composition with $\delta^{17}O \approx \delta^{18}O \approx -50$ ‰, reflected in the composition of CAIs and confirmed by results of the GENESIS mission.

The existence of a $^{17}O$- and $^{18}O$-rich $H_2O$ reservoir is suggested by the observation of high $\Delta'^{17}O$ in the most oxidized phases (magnetite) in Semarkona (Choi et al. 1998). These are suggested to have formed by reaction of Fe metal or troilite with $^{16}O$-poor water. Sakamoto et al. (2007) reported oxidized phases in chondrites with $\Delta'^{17}O$ as high as +80 ‰. This is within the range of predicted $H_2O$ (ice) in the model of Lyons and Young (2005) with an enhanced UV flux of only 5× (relative to the local interstellar medium). Organic material with even higher $\Delta'^{17}O$ has been analyzed by Hashizume et al. 2011. All these observations are in agreement with the existence of $^{16}O$-poor $H_2O$. On the other end, CO formed the counterbalancing $^{16}O$-rich reservoir that had been lost. If the bulk SS is at $\delta^{17}O = \delta^{18}O = -50$ ‰, CO is expected to have had lower $\delta^{17}O$ and $\delta^{18}O$ values. So far, only three objects have been analyzed that may reflect the composition of the lost $^{16}O$-rich CO reservoir: a chondrule from the CH3 Acfer 214 carbonaceous chondrite (Kobayashi et al. 2003), and a two CAIs from the CH/CB-like Isheyevo carbonaceous chondrite (Gounelle et al. 2009).

Despite the success of these models, debate remains about the exact origin of O isotope anomalies in SS materials. For instance, a line of arguments was presented by Thiemens and Lin (2021) that criticized the self-shielding model in the form suggested by Clayton (2002) and Lyons and Young (2005). Also, experiments were not able to reproduce the effects predicted by the self-shielding model (Chakraborty et al. 2008). Some observations are also surprising when seen through the lens of the self-shielding hypothesis. For instance, apart from magnetite in some unequilibrated OC chondrites, no chondrite silicates and oxides have been observed that plot towards the $^{16}O$-poor self-shielding-produced $H_2O$ with $\Delta'^{17}O > 7$‰ (Choi et al. 1998; Doyle et al. 2015). While the high $\Delta'^{17}O$ of magnetite in unequilibrated OCs is thought to be inherited from the aqueous fluids on the parent bodies, the corresponding alteration products (fayalite, magnetite) from some CCs plot below the TFL (Doyle et al. 2015). This suggests that the water on these CC parent bodies had lower $\Delta'^{17}O$ than the water in the unequilibrated OC parent body, which is opposite to what would be expected since CC parent bodies are thought to form at greater heliocentric distances.

*(ii) Systematics in bulk meteorite data*

Irrespective of the origin of O isotope anomalies, in a number of seminal papers, Clayton and his group explored the composition of bulk meteorites (Clayton 2008; Clayton 2003; Clayton 1993; Clayton et al. 1991; Clayton et al. 1976; Clayton and Mayeda 1999; Clayton and Mayeda 1996; Clayton and Mayeda 1984; Clayton and Mayeda 1983). A number of laboratories now analyze triple O isotopes at high-precision and the amount of data has increased substantially. We therefore summarize bulk rock O isotope data compiled from the Meteoritical Bulletin database (https://www.lpi.usra.edu/meteor/) and display the data in Fig. 7.

The main chondrite groups span a large range in $\Delta'^{17}O$ from -5 to +3 ‰, *i.e.* spanning some 8000 ppm (Fig. 7a). The observed bulk-rock variations are large in comparison to the analytical resolution in the range of <10 ppm. With exception of the aqueously altered CI chondrites, all carbonaceous chondrites have in common that they are enriched in $^{16}$O compared to the non-carbonaceous chondrites, the Earth and most differentiated meteorites, most likely due to the incorporation of $^{16}$O-rich refractory inclusions. The achondrites, which source from parent bodies that have experienced melting and lost the primary phases and textures preserved in chondrites, mostly fall in a narrower range of $-0.5 \leq \Delta'^{17}O \leq +1.5$ ‰ (Fig. 7b). Only the ureilites, the Eagle Station group pallasites, the IVB iron meteorites, and some ungrouped meteorites are an exception here and fall in the field of carbonaceous chondrites along the CCAM. Despite their O isotope similarity, ureilites belong to the NC reservoir (Fig. 7c), while the other meteorites have been linked to the CC reservoir. Finally, a large set of meteorites ($N$ = 94) from the Meteoritic Bulletin Database have been classified as "ungrouped achondrites". The data are shown in Fig. 7c. They suggest additional distinct parent bodies in the fields typical of CC and NC achondrites, as well as underexplored links between meteorites sampling different reservoirs within the same parent body. The data demonstrate that the collection of large datasets, *e.g.*, in the Meteoritical Bulletin Database, can be of high value for identifying distinct reservoirs and genetic relationships.

Despite the historical significance of, and fundamental discoveries brought on by, O isotope anomalies in meteorites, their interpretation is complicated by the fact that they are influenced by multiple processes (*e.g.*, gas-melt interactions, water interactions). Hereafter, we will therefore focus on the more recent insights gained into genetic relationships from nucleosynthetic anomalies, which only trace mixing.

### *3.3 Nucleosynthetic anomalies*
#### *(i) Data and systematics*

Present in their purest form in presolar grains, nucleosynthetic isotope anomalies are also found, diluted, in chondrite components like refractory inclusions or chondrules, and remain detectable even at the bulk meteorite- and planetary-scale. The anomalies observed in SS materials indicate that different types of stellar sources contributed material to the Sun's parent molecular cloud. How this cloud-scale heterogeneity relates to the isotope anomalies of chondrite components and planetary bodies forming in the protosolar disk, and hence, what we can learn from these anomalies about early SS evolution, remains a matter of intense debate. To guide this debate, we present here an overview of the most up-to-date dataset of isotope anomalies in SS materials. These anomalies exhibit systematic trends in multi-element isotope space, reflecting a handful of key nucleosynthetic pathways in stars and supernovae, and forming the basis for any spatial and temporal interpretations regarding the dynamics of SS formation (*Sections 4 and 5*).

Figure 8 summarizes the collective effort of the community over the past two decades and shows the distribution of isotope anomalies among bulk meteoritic and planetary materials in multi-element isotope space. The elements shown in Fig. 8 were chosen for illustrative purposes and cover a range of nucleosynthetic origins and geo-/cosmochemical behaviors. As is evident, the most salient feature of this dataset is a fundamental isotopic difference among bulk SS materials (red vs. blue symbols), referred to as the NC-CC dichotomy (*Section 3.1*). This clustering of the data is most pronounced in spaces were the anomalies are large (*e.g.*, $\varepsilon^{54}$Cr vs $\varepsilon^{50}$Ti), but is also visible when the anomalies are small (*e.g.*, $\varepsilon^{54}$Fe vs $\varepsilon^{64}$Ni). The NC group contains enstatite (EC), ordinary (OC), Rumuruti, and Kakangari chondrites, while the CC group contains the carbonaceous chondrites (*e.g.*, CI, CV, CO, CM, CR). Importantly, both groups contain iron meteorites and stony achondrites, sampling differentiated planetesimals that accreted early in the protosolar disk relative to chondrites.

The Earth, Moon, and Mars belong to the NC group, with the isotopic composition of the Bulk Silicate Earth (BSE) being closest to that of ECs, and that of the Bulk Silicate Mars (BSM) being typically intermediate to ECs and OCs. As such, it is commonly invoked that the NC group represents relatively volatile-poor, inner SS materials, while the CC group reflects materials richer in volatiles accreted at greater heliocentric distance (*e.g.,* Hilton et al. 2022; Kleine et al. 2020; Spitzer et al. 2021).

The separation of materials into two nebular reservoirs was first proposed based on the $\varepsilon^{54}$Cr–$\Delta^{17}$O systematics of differentiated NC bodies, OCs, and CC chondrites (Trinquier et al. 2007). A few years later, Warren (2011) compiled available isotope anomaly data for Cr, Ti, Ni, and O and showed that meteoritic materials defined two distinct clusters in isotope-isotope plots for these elements. Importantly, all meteorites consistently fall into one of the two isotopic clusters regardless of the combination of elements considered, indicating this bimodality is a ubiquitous feature among SS bodies. The NC–CC dichotomy has since been observed for numerous other elements with different geochemical affinities (Table 1). These include the siderophiles Mo, Ru, Fe, W, Pd, and Pt, and lithophiles Ca, Si, Sr, Zr, Ba, Sm, and Nd. All these elements are refractory to moderately refractory, meaning that they condense at relatively high temperatures (>1250 K) from a gas of solar composition (Lodders 2003). In contrast, nucleosynthetic isotope anomalies in volatile elements – with half-mass condensation temperatures ($T_C$) below those of the major rock-forming elements Mg, Si, and Fe (<1250 K) – were long believed to be absent due to extensive homogenization within the disk (Vollstaedt et al. 2020). Recent studies have, however, identified nucleosynthetic isotope anomalies for the lithophile and moderately volatile elements Zn ($T_C \sim 704$ K) and K ($T_C \sim 993$ K) (Martins et al. 2023; Nie et al. 2023; Savage et al. 2022; Steller et al. 2022) providing an additional valuable tool for constraining the provenance of Earth's (and Mars') volatiles. The fact that the dichotomy persists across elements with different cosmochemical and geochemical behaviors is an important clue to understanding its origin (*Section 4*).

Returning to Fig. 8, the *inter*-reservoir trend between NC and CC bodies (*i.e.,* the dichotomy) can be described as an enrichment of supernova produced (NSE and *r*-process) isotopes in CC bodies relative to NC bodies. The NC-CC dichotomy, phrased more specifically, is the systematic offset of CC bodies from NC bodies towards CAIs in isotope-isotope space. Two families of models have been proposed to explain the origin of the dichotomy, which are discussed in detail in *Section 4*. Before diving into this interpretative discussion, we continue with a description of the structure of the data. Closer examination of Fig. 8 also reveals finer structure beyond the NC-CC dichotomy: *i.e., intra*-reservoir trends. In particular, NC bodies define a linear array consistent across multi-element isotope space, with the BSE as an endmember. Depending on the elements considered, the NC array points either towards the CC group (and CAIs), or away from them. In the former case, the BSE is positioned either closest to, or farthest away from, the CC bodies. An *intra*-reservoir trend in the CC group is only clearly apparent in some isotope spaces (*e.g.,* $\varepsilon^{50}$Ti-$\varepsilon^{54}$Cr), and the spread in the data mostly correlates with the relative abundance of CAIs, chondrules, and matrix in CC chondrites (Alexander 2019; Hellmann et al. 2023; Hellmann et al. 2020).

The exact morphology of the ε-ε plots (*i.e.,* 2D cross sections of multi-element isotope space) reflects the nucleosynthetic origin of the isotope anomaly being considered. These can be separated into 5 sub-categories, corresponding to the combinations of the different nucleosynthetic pathways responsible for these anomalies:

(a) When two isotopes whose anomalies result from NSE contributions (*e.g.,* $^{50}$Ti, $^{48}$Ca, $^{54}$Cr, $^{66}$Zn; Fe and Ni excluded, see below) are plotted against each other (Fig. 8a), the NC trend points towards the CCs, with the BSE endmember closest to the CCs.

(b) When anomalies attributed to NSE are plotted against those of Fe or Ni (Fig. 8b), the BSE remains closest to the CCs, but the NC trend points away from them, and the CI chondrites separate from the CC cluster.

(c) When anomalies attributed to NSE are plotted against those of elements beyond the Fe-peak (*i.e.,* reflecting excesses in *s*- and *r*-process nuclides; Mo, Zr, Ru, Ba) (Fig. 8c), the data distribution is akin to (b) but CI chondrites are not clearly distinct from the other CCs. Instead, CI chondrites appear to be an endmember of the CC group (as it does for anomalies of all elements except Fe and Ni).

(d) When plotting *s*- and *r*-process anomalies against those in Fe or Ni (Fig. 8d), the NC trend points towards the CCs as in (a), but the BSE sits farthest from the CCs. The CI chondrites are separated from the rest of the CCs.

(e) Finally, when *s*- and *r*-process anomalies are plotted against each other, a data distribution akin to that in (d) is retrieved, but with CI chondrites seemingly in the CC group (Fig. 8e).

At first glance, two elements do not seem to conform to the general trends described above (Fig. 9). We consider them in more detail below.

*Silicon:* Silicon isotope anomalies ($\mu^{30}$Si) were recently reported in bulk SS materials by Onyett et al. (2023). In contrast to other elements, NC and CC chondrites have overlapping anomalies, and a dichotomy is seen between chondrites and NC achondrites instead. Taken at face value, these observations suggest a temporal evolution of the inner disk's Si isotopic composition from NC achondrite- to CI-like. However, Dauphas et al. (2024) pointed out that the elevated $\mu^{30}$Si values in enstatite chondrites may be arising from an improper correction of mass-dependent fractionation. Using the equilibrium law (relevant to fractionation at high temperatures) instead of the exponential law would result in a $^{30}$Si anomaly of enstatite chondrites within error of the BSE. This would align the Si isotope data more closely with the general trends described in the above section (Fig. 9a). With only one study to date, future studies are needed to resolve this debate.

*Strontium:* With only four isotopes ($^{84}$Sr, $^{86}$Sr, $^{87}$Sr, and $^{88}$Sr), and one ($^{87}$Sr) typically produced overwhelmingly by decay of $^{87}$Rb, it has been famously difficult to identify the nucleosynthetic origin of strontium (Sr) isotope anomalies. After normalization to $^{86}$Sr and $^{88}$Sr, anomalies are only apparent in $^{84}$Sr. Whether these represent true variations in $^{84}$Sr abundance (*i.e.*, a result of the *p*-process) or variations in abundances of the normalizing isotopes ($^{86}$Sr, *s*-process; $^{88}$Sr: *s*- and *r*-process) has been a long standing question (Burkhardt et al. 2019; Charlier et al. 2021; Charlier et al. 2019; Fukai and Yokoyama 2019; Hans et al. 2013; Moynier et al. 2012; Papanastassiou and Wasserburg 1978; Paton et al. 2013; Schneider et al. 2023; Yokoyama et al. 2015). Most relevant to our discussion is the systematics observed in bulk meteorites and planetary materials. As for other elements, CC meteorites have Sr anomalies that are generally offset from NC materials towards those of CAIs. Unlike most other elements, however, NC materials have indistinguishable composition and some CC chondrites have anomalies similar to NC materials (Charlier et al. 2017; Schneider et al. 2023) (Fig. 9b). These apparent departures from the general trends described above (Fig 8) were recently proposed to reflect variable *s*- and *(r, p)*-process contributions (Schneider et al. 2023). In NC meteorites, these variations are correlated, resulting in $^{84}$Sr excesses and deficits of similar magnitude that effectively cancel each other out. In CC meteorites, the presence of CAIs (which have elevated $^{84}$Sr anomalies) leads to the apparent $^{84}$Sr excess. In the absence of additional Sr isotopes to unambiguously quantify *s*-, *r*- and *p*-process variations in SS materials, the proposal of Schneider et al. (2023) leverages the isotope-isotope trends seen in other elements to explain the seemingly different Sr systematics.

*(ii) Constraints from inferred accretion ages*

Armed solely with the dataset of nucleosynthetic anomalies (Fig. 8), it is impossible to tell if the structure in the data described above reflects a temporal and/or spatial evolution of the early SS. In particular, the NC-CC dichotomy could reflect (i) the simultaneous existence of two spatially separated reservoirs, or (ii) a temporal evolution in the isotopic composition of the disk as whole (Warren 2011). Distinguishing between these two hypotheses requires coupling the genetic information from nucleosynthetic isotope anomalies in NC and CC meteorites to the chronological information of the accretion age of their parent bodies.

Accretion ages of meteorite parent bodies are inferred through a combination of isotopic analyses and thermal models of internal heating from $^{26}$Al decay. For stony achondrites and iron meteorites, thermal modeling is anchored by Hf-W model ages of core segregation: ~1-2 Myr and ~3 Myr after CAI condensation for NC and CC iron meteorite parent bodies, respectively (Spitzer et al. 2021). For chondrites, thermal models are informed by either their cooling histories following thermal metamorphism, as deduced from Pb-Pb (*e.g.,* Blackburn et al. 2017) and Hf-W systematics (*e.g.,* Hellmann et al. 2019), or radiometric dating of aqueous alteration (*i.e.,* via carbonates) by the Mn-Cr system (*e.g.,* Jogo et al. 2017). Combined, these techniques indicate that planetesimal formation took place as early as several $10^5$ years after CAIs, and lasted for ~4 Myr (Budde et al. 2018; Kruijer et al. 2020; Neumann et al. 2023; Spitzer et al. 2021; Sugiura and Fujiya 2014). This is roughly consistent with the timing of disk dissipation as inferred from paleomagnetic studies (Borlina et al. 2022; Wang et al. 2017), and the extinct radionuclide (*e.g.,* Al-Mg) systematics of chondrules (Dauphas and Chaussidon 2011). These records imply an inside-out gradual disk dispersal that limited planetesimal accretion in the inner disk to about 2 Myr after CAIs, while in the outer disk parent-body formation continued until about 4.5 Myr after CAIs. The bodies forming within ~1.5 Myr after CAIs differentiated due to the presence of $^{26}$Al and are sampled by iron meteorites and stony achondrites, while chondrites sample bodies assembled from about 2 Myr onwards, when $^{26}$Al decayed to levels insufficient to induce melting and differentiation (*e.g.*, Spitzer et al. 2021).

Based on an apparent linear relationship between $\varepsilon^{54}$Cr and the time of accretion for NC stony meteorites and CC chondrites, Sugiura and Fujiya (2014) supported the hypothesis of a temporal evolution of the isotopic composition of the disk. This conclusion, however, was based on a limited dataset, devoid of CC stony achondrites (only a few tens of which have been identified so far; Yap and Tissot 2023), and iron meteorites. In Figure 10, we plot the $\varepsilon^{54}$Cr of both stony and iron meteorites from NC and CC groups against their inferred time of accretion. The data reveals an overlap between NC and CC parent body accretion ages, testifying to the fact that the NC and CC reservoirs co-existed in separate regions of the early SS, and that there was no simple evolution of the isotopic composition within or between the reservoirs over time. The same observation can be made using nucleosynthetic anomalies of other elements (*e.g.,* Fe, Ni, Mo, Ru), which, by virtue of their siderophilic affinities, are present in both stony and iron meteorites.

## 4. Origin of large-scale isotopic heterogeneity in the early SS

Two families of models have been proposed to explain the origin of the NC-CC dichotomy (Fig. 11). In the first (Fig. 11a-b), the dichotomy reflects the inheritance of isotopic heterogeneity initially present in the molecular cloud, and CC bodies constitute mixtures of isotopically NC and CAI-like materials, representing late and early infall, respectively (*e.g.,* Burkhardt et al. 2019; Nanne et al. 2019; Yap and Tissot 2023). In the second (Fig. 11c-d), the presolar carriers were homogeneously distributed within the cloud, and isotopic heterogeneity arose from sorting of presolar carriers through thermal and physical processing within the disk (*e.g.,* Burkhardt et al. 2012; Dauphas et al. 2010; Davis et al. 2018; Trinquier et al. 2009).

### *4.1. Tracing the molecular cloud infall*

The *intra-* and *inter-*reservoir trends presented in Section 3.3 may reflect a plethora of superimposed nebular processes. Nonetheless, their persistence across elements with a wide range of geo- and cosmo-chemical behavior suggests they are primarily governed by mixing between several isotopically distinct nebular components inherited from the parent molecular cloud (Fig. 11a-b). Magnetohydrodynamic models of protostellar accretion indicate significant variance in the source of infalling material within collapsing molecular clouds, and associated infall timescales (*e.g.,* Kuffmeier et al. 2023). These astrophysical considerations lend credence to models that suggest the spatial order of isotopic heterogeneity in the nascent disk derives from a chronological order of distinct infalling materials (Burkhardt et al. 2021; Burkhardt et al. 2019; Jacquet et al. 2019; Kleine et al. 2020; Nanne et al. 2019; Van Kooten et al. 2016; Yap and Tissot 2023). Here, we review the key aspects and recent developments of such models, as well as the enticing avenues for future work they bring forth.

#### *(i) Inheritance models*

What we henceforth denote as *inheritance models* rely on the idea that the apparent trend in multi-element isotope space from NC to CC to CAI compositions is the result of mixing of isotopically distinct endmembers (Fig. 8). This trend was initially observed for Ti isotopes, and the anomalous Ti isotope composition of carbonaceous chondrites was interpreted as reflecting variable mixtures of their isotopically distinct, major constituents: matrix, chondrules, and CAIs (Niemeyer 1988a; Niemeyer 1988b; Niemeyer and Lugmair 1984), which themselves formed from precursor materials that retained a "cosmic chemical memory", *i.e.*, isotopic heterogeneities inherited from the molecular cloud (Clayton 1982). In his seminal work, Niemeyer (1988a) further deduced that CI chondrites, which are taken as a proxy for the chemical composition of the Sun for all but the volatile elements, may not be representative of the *isotopic* composition of the bulk SS, and that mass-balance dictates that "*our present collection of [bulk] SS samples is missing a large reservoir that is more enhanced in $^{50}Ti$ (e.g., the Sun)*". All isotope anomaly data obtained since then for carbonaceous chondrites and their components may be interpreted along the premises laid out by Niemeyer (1988a). By now, it is evident that the offset of CC meteorites from the NC meteorites towards the isotopic composition of CAIs is a global feature seen for all elements showing isotope anomalies (Fig. 8). Importantly, this includes not only elements that are enriched in CAIs (*e.g.*, Ca, Ti, Sr), but also elements that are depleted in CAIs (*e.g.*, Cr, Ni). Hence, while CAIs *sensu stricto* play a role in setting the isotopic composition of carbonaceous chondrites for refractory elements, they do not contribute significantly to the anomalies in less-refractory elements. This observation led to the postulation that material with an isotopic composition like CAIs, but a less-refractory chemistry, must be present within carbonaceous chondrites. In other words, CAIs are the refractory part of an isotopically distinct reservoir in the disk, and both refractory and non-refractory products forming from this reservoir must have contributed to the makeup and isotopic composition of carbonaceous chondrites (Burkhardt et al. 2019; Nanne et al. 2019; Yap and Tissot 2023). The existence of such an Inclusion-like Chondritic (IC) reservoir (Burkhardt et al. 2019) is supported by (i) the finding of common Ti, Cr, and O isotope anomalies in CAIs and the less-refractory AOAs from CV chondrites (Jansen et al. 2024; Torrano et al. 2024), and (ii) the variable Ti and Cr isotopic composition of individual chondrules from CV and CK carbonaceous chondrites, ranging from NC to the hypothetical IC compositions (Fig. 12) (Gerber et al. 2017; Olsen et al. 2016; Schneider et al. 2020; Williams et al. 2020; Zhu et al. 2019). Hence, the

CC reservoir can be understood as a mixture of NC and IC materials (CC=NC+IC), whose distinct isotopic compositions may be directly inherited from the molecular cloud through infall processes.

CAIs are the SS's oldest solids and formed at high temperatures, most likely close to the young Sun, but are now most abundant in bodies forming later in the outer disk (*i.e.*, carbonaceous chondrites and comets; Joswiak et al. 2017; Krot et al. 2019; McKeegan et al. 2006). This requires the transport of CAIs to the outer disk, a process that is naturally explained by (i) radial disk expansion accompanying material infall close to the protostar (*i.e.*, viscous spreading; expansion of the centrifugal radius), and/or (ii) ballistic mechanisms (*i.e.*, disk winds and/or stellar outflows) (*e.g.*, Desch et al. 2018; Hueso and Guillot 2005; Morbidelli et al. 2024; Scott et al. 2018; Yang and Ciesla 2012 & references therein). The virtual absence of CAIs in NC meteorites suggests CAI transport (either to the outer disk, or into the Sun due to radial drift) was quantitative, and implies that material falling in early and close to the protosun became enriched in the outer disk, while late infalling material dominated the inner disk (Nanne et al. 2019).

By combining infall and disk evolution models with the isotopic signatures of meteorites, a framework describing the early evolution of the SS emerged that, despite its simplicity, can explain many observables of the meteoritic record (Burkhardt et al. 2019; Jacquet et al. 2019; Nanne et al. 2019; Yap and Tissot 2023). In essence, this model equates the IC reservoir with early infalling cloud material, and the NC reservoir with late infalling cloud material. Hence, it implies that many of the overarching isotopic variations among planetary materials in the SS ultimately date back to the infall and disk building phase and are inherited from an isotopically heterogeneous molecular cloud core. A cartoon of the model in its simplest form is provided in Fig. 11a-b, and its isotopic implications are shown on Fig 13.

During the initial stages of collapse, infalling material is IC in composition and limited to the vicinity of the protosun, where it was thermally processed before being transported outwards by viscous spreading. As such, the nascent disk and a significant portion of our Sun would have been CAI-like in isotopic composition: *i.e.*, enriched in NSE and *r-process* nuclides (see quote from Niemeyer (1988a) above). With time, as the disk expanded, the composition of the infalling material transitioned to be more NC-like. Since most of the infalling mass is assumed to be deposited close to the Sun, the later NC dust dilutes and overprints IC isotopic signatures in the inner disk. As infall dwindles, the disk thus formed is characterized by a gradient from NC material in the inner disk to IC material in the outer disk, with a mixture of NC and IC materials (*i.e.,* CC material) in-between. To a certain extent, this primordial disk gradient must have been retained during planetesimal formation (see Section 3.3.iii). Most likely, this happened via pressure bumps in the disk (*e.g.,* induced by Jupiter's formation at the water-ice sublimation line), which blocked the passage of inward drifting IC materials and triggered NC and CC planetesimal accretion (*e.g.*, Hellmann et al. 2023; Kruijer et al. 2017; Lichtenberg et al. 2021; Morbidelli et al. 2022).

A number of studies explored the permittable parameter space of the infall models with respect to the meteoritic constraints in more detail. They indicate that the angular momentum of the infalling material must have been rather low in order to constrain to centrifugal radius close to the star, while the initial viscosity of the disk must have been high ($\alpha_0 > 0.01$), and the infall time short ($<5\times10^5$ yr) to allow for high rates of viscous spreading with CAI outward transport and massive outer disk buildup (Jongejan et al. 2023; Marschall and Morbidelli 2023; Woitke et al. 2024). Magnetic braking and magnetohydrodynamic effects might allow for these constraints to be met, but thus far these effects have not been included in the parameter studies directly. Nevertheless, these studies all agree that infall and viscous spreading likely played a key role in setting the initial conditions for the SS. Hence, while the details remain to be ironed out, and the model setup is certainly oversimplified (*e.g.*, the IC material spherically contained in the surrounding NC material layer), it must be highlighted that the very existence of a self-consistent model explaining the meteoritic evidence within an astronomical framework is a success in itself.

*(ii) CI as chemical but not isotopic proxies of primitive SS dust*

In the inheritance model, the CC meteorites, including the CI chondrites, represent variable mixtures of NC and IC materials. This is of interest since the chemical composition of CI chondrites is taken as representative of the non-volatile element composition of the Sun (Asplund et al. 2009; Lodders 2003; Palme et al. 2014), and in chemical and isotopic mixing models of carbonaceous chondrites, CI's are used as representative of a matrix endmember component (*e.g.*, Alexander 2019; Hellmann et al. 2020; Nie et al. 2021). Thus, CI chondrites are generally thought to be good proxies for the unprocessed primitive dust in the solar accretion disk, and also are used in this way in models assigning the origin of nucleosynthetic anomalies in CAIs and bulk meteorites to differential dust processing in the disk (see *Section 4.2*). However, there is growing evidence that CI chondrites themselves are mixtures, lending credence to the idea that they formed from NC and IC materials. For instance, it was found that compared to NC bodies, CI chondrites like other CC chondrites exhibit an excess in thulium (Tm) relative to neighboring rare earth elements, indicating the presence of CAIs in their precursor materials (Barrat et al. 2016; Dauphas and Pourmand 2015). Likewise, based on a trend between $\varepsilon^{50}$Ti anomalies and Ti/Si elemental ratios among chondrites, it was inferred that the precursor materials of CI chondrites might have contained up to 3 % of refractory CAI-like dust (Bryson and Brennecka 2021; Burkhardt et al. 2019). The presence of IC material is further supported by reports of relict CAIs (up to ~ 0.5 atom. %; Frank et al. 2023) and AOA-like olivines (Kawasaki et al. 2022; Morin et al. 2022) in CI chondrites. Together with the finding of chondrule-like olivines (Kawasaki et al. 2022), these observations point towards a bi-modal origin of refractory dust in CI chondrites: some being $^{16}$O-rich like CAIs, AOAs, and the Sun, and some being $^{16}$O-depleted like the NC materials. We note that AOAs have about chondritic relative abundances of refractory elements (Jansen et al. 2024; Torrano et al. 2024), such that a significant amount of AOAs could be accommodated in CI chondrites without changing their overall solar-like bulk composition. Furthermore, it has been suggested that the uniquely distinct Fe and Ni isotopic compositions of CI compared to other CC chondrites reflect fractionation of a metal component among their precursor materials, possibly because their formation was triggered by photoevaporation of the gas towards the end of the disk's lifetime (Spitzer et al. 2024). Hence, while CI chondrites have a chemical composition similar to the Sun, they nonetheless appear to be mixtures of isotopically distinct source materials. It is therefore possible that the *isotope* composition of CI's is not representative of the average unprocessed primitive dust from which our SS and the solar accretion disk formed. In fact, a growing numbers of studies have documented the existence of various chemically CI-like materials with different nucleosynthetic compositions, such as CI-like dark clasts in different host chondrites (Goodrich et al. 2021; Van Kooten et al. 2024; Van Kooten et al. 2017b).

*(iii) Origin of the intra-reservoir NC trend: missing end-member*

Aside from the NC-CC-IC offset, the most salient feature in the isotope anomaly data of the bulk planetesimal and planetary bodies is the global correlation of the anomalies in the NC reservoir (Fig. 8). These correlations are present amongst elements with different cosmochemical (refractory to moderately volatile) and geochemical (lithophile to siderophile) behavior, and among isotopes of different nucleosynthetic origin (stellar and explosive nucleosynthesis in different types of low- and high-mass stars). Furthermore, the anomalies of different nucleosynthetic heritage are not only correlated among different elements, but also among different isotopes of the same element. For instance, correlated variations in $\varepsilon^{46}$Ti and $\varepsilon^{50}$Ti are associated with anomalies originating in Type Ia and Type II supernovae, respectively (*e.g.*, Davis et al. 2018; Zhang et al. 2011). Likewise, variations in $\varepsilon^{94}$Mo and $\varepsilon^{95}$Mo among NC meteorites were found to contain correlated contribution of *s*- and *r*-process nuclides (Spitzer et al. 2020), as does the $\varepsilon^{84}$Sr signature of NC bodies (Schneider et al. 2023). Moreover, no trend is apparent between the formation age of NC bodies and their isotopic composition (Fig. 10). Together, these features of the NC trend imply that the global isotopic variations are the result of a two component mixture, where one mixing endmember is enriched in *s*-process and NSE and supernova-associated nuclides relative to the composition of the BSE (*e.g.*, positive $\varepsilon^{30}$Si, $\varepsilon^{48}$Ca, $\varepsilon^{50}$Ti, $\varepsilon^{54}$Cr, $\varepsilon^{64}$Ni, $\varepsilon^{66}$Zn, $\varepsilon^{100}$Ru, and negative $\varepsilon^{54}$Fe, $\varepsilon^{96}$Zr, $\varepsilon^{94}$Mo) and the other is depleted in these nuclides. The exact nature and composition of the mixing endmembers remain vague, but the mixing trend must have been established early, and is unlikely to be caused by the differential processing of individual carriers, or by fractionation due to specific element properties, as these processes cannot easily explain the *global* nature of the observed correlations. It has thus been suggested that the NC isotopic anomaly trend, like the NC-CC-IC offset, is related to heterogeneous infall and represents a gradient in the inner disk (*e.g.*, Spitzer et al. 2020). In such a scenario the very end of infall would represent the lost inner SS mixing endmember that balances any contribution of known meteoritic material to the forming Earth, and which is expected to dominate the composition of the disk sunwards of Earth's orbit (Burkhardt et al. 2021; Yap and Tissot 2023).

In summary, the heterogeneous accretion and selective processing of NC and IC materials in different nebular environments, combined with the heterogeneous distribution of the resulting nebular products in the disk can readily account for the observed isotopic and elemental variations between, and to a large extent also within, the NC and CC reservoirs (Burkhardt et al. 2019; Hellmann et al. 2023; Jacquet et al. 2019; Nanne et al. 2019; Yap and Tissot 2023). As such, much of the compositional diversity of meteorites seem to have its origin in the very initial stages of disk building and evolution.

*(iv) Implications for short-lived radionuclides*

The interpretation of the isotopic anomaly record as an inherited signature of heterogeneous infall is in line with recent astronomical observations of filamentary infall and disk substructures, and the thereby initiated change in the perception of how and when material can be delivered to planet-forming disks (*Section 2*). Streamers and other observed infall phenomena provide compelling mechanisms for delivering isotopically heterogeneous cloud material to the evolving protosolar disk. Such heterogeneous infall not only can explain the overarching NC/CC dichotomy, but also a range of other meteoritic anomaly signatures. For instance, some rare CAIs seem to have formed with no or much lower initial abundance of the short-lived nuclide $^{26}$Al than normal CAIs. These include the so-called PLACs (platy hibonite crystals, Ireland 1988), and hibonite or corundum dominated FUN CAIs (Fractionation and Unidentified Nuclear effects; Desch et al. 2023; Krot et al. 2014 and references therein). These $^{26}$Al-poor CAIs often also have variable and large nucleosynthetic anomalies and are generally interpreted to have formed before the addition of $^{26}$Al to the SS (*e.g.*, Kööp et al. 2018, however also see Desch et al. 2023 for an alternative interpretation). Within the IC-NC inheritance scenario this would imply that the IC infall forming the normal CAIs was preceded by an even earlier infall of $^{26}$Al-free cloud material (*e.g.*, Brennecka et al. 2020). However, given the observation of streamers at different stages of disk evolution, an alternative interpretation of the $^{26}$Al-free CAIs could be their formation at a later time from material zapped onto the disk by a short-lived streamer sampling cloud material not polluted by freshly synthesized $^{26}$Al. To what extent such transient heterogeneities in the accreting materials affected the bulk isotopic evolution of the disk, and in particular the debated distribution of $^{26}$Al among forming planetary bodies (*e.g.*, Bollard et al. 2019; Budde et al. 2018; Desch et al. 2023; Jacobsen et al. 2008; Kita et al. 2013; Krestianinov et al. 2023; Larsen et al. 2020; Larsen et al. 2011; Schiller et al. 2015; Van Kooten et al. 2016; Wasserburg et al. 2012) remains to be investigated. In any case, the possibility of short-lived streamers should be taken into account in any future assessment of isotopic variability among solar system materials.

### *4.2. Thermal processing*

A second family of models, which we henceforth denote as *unmixing models* (Fig. 11c-d), have been proposed to explain the isotopic heterogeneity of SS materials. These models rely on the idea that presolar carriers were initially homogeneously distributed within the cloud of CI chondritic chemical and isotopic composition, and that the sorting of the carriers through thermal and physical processes in the disk led to the isotopic structure seen today (*e.g.,* Burkhardt et al. 2012; Dauphas et al. 2010; Davis et al. 2018; Trinquier et al. 2009).

Recent sample return missions from the asteroids Ryugu and Bennu have shown that CI chondrites likely represent an abundant reservoir in the outer SS (Lauretta et al. 2024). A similar idea had previously been suggested by models of CC matrix compositions (Alexander 2019; Hellmann et al. 2020; Nie et al. 2021), as well as empirical data of CC matrix (Van Kooten et al. 2021; Van Kooten and Moynier 2019; Van Kooten et al. 2019; Zanda et al. 2018). In contrast to inheritance models that invoke the hypothetical IC component as a large missing reservoir of the outer SS, unmixing models thus build on the proposed ubiquity of CI-like material in the outer SS. It is then the thermal processing of this CI-like material in the early protoplanetary disk that would lead to its unmixing into two dust populations.

Physico-chemical gradients are inherent to accretion disks and the thermal processing of dust is thought to be a generic process in at least the inner disk regions (*e.g.*, Colmenares et al. 2024; Henning and Semenov 2013). Given the variable properties of different (presolar) minerals, and the variable cosmo- and geochemical behavior of different elements, it is thus conceivable that the selective processing of presolar solids in the protosolar disk induced a significant fraction of the isotopic anomalies seen among planetary materials. Indeed, many of the isotope anomalies at the bulk meteorite scale have been assigned to disk processing in their discovery publications. For instance, variable *s*-process Mo deficits among meteorites were assigned to the selective processing of presolar SiC (Dauphas et al. 2002b), as was the correlation between Mo and Ru isotope anomalies (Dauphas et al. 2004), and Nd variations among carbonaceous chondrites (Carlson et al. 2007). Likewise, the observation of Ni isotope anomalies among meteorites was suggested to be caused by size-sorting of presolar dust grains during transport within the disk (Regelous et al. 2008).

Based on correlated anomalies in $\varepsilon^{46}$Ti, $\varepsilon^{50}$Ti, and $\varepsilon^{54}$Cr among planetary materials Trinquier et al. (2009) proposed a thermal unmixing model of CI-like dust that has been frequently adopted in later works (*e.g.*, Ek et al. 2020; Larsen et al. 2011; Paton et al. 2013; Poole et al. 2017; Schiller et al. 2015). In essence, this model suggests CI chondrites represent the precursor dust from which the protosun and the surrounding disk accreted, and that the inter-reservoir isotopic anomaly trend of NC, CC, and CAI materials can be explained by unmixing of this dust by thermal processing (Fig. 11 & 13). This would imply that even though $^{46}$Ti, and $^{50}$Ti and $^{54}$Cr have different nucleosynthetic origins and their carriers may not have a similar mineralogy, they must have similar physicochemical characteristics that allow them to be thermally processed together. Trinquier et al. (2009) suggested that, since chondritic matrix shows evidence of thermal processing by loss of moderately volatile elements (Bland et al. 2005), a similar process must have destroyed the thermally labile $^{54}$Cr, $^{46}$Ti, and $^{50}$Ti carriers. In more detail, the thermal processing model invokes the sublimation of thermally labile over thermally robust isotopically anomalous material. The gas phase forming during the thermal processing of labile carriers is then progressively enriched in supernova-derived nuclides ($^{54}$Cr, $^{50}$Ti, $^{26}$Al, $^{48}$Ca, $^{88}$Sr) and eventually forms CAIs and AOAs with this signature by condensation, whereas the residual material is depleted in these isotopes relative to the CI starting composition and ends up in NC bodies. Alternatively, it has been proposed that the thermally labile carrier is represented by ISM-derived dust or icy and organic-rich mantles around silicate dust (Ek et al. 2020). In this way, the isotopic divide may also relate to the petrological divide between water-poor NC and water-rich CC bodies.

In any case, although thermal processing of dust in protoplanetary disks seems unavoidable, a general concern of the thermal processing models is that it remains unclear how they work in detail. In particular, they have to overcome the challenge that the carriers of different types of anomalies and different elements likely respond differently to thermal processing; yet, at the bulk meteorites scale, there are consistent isotope anomaly correlations between elements with different nucleosynthetic heritage and different cosmochemical and geochemical properties (Fig. 8). A profound hurdle on the way to a better understanding of the relevance of thermal processing for the generation of bulk parent body anomalies is the lack of knowledge about the properties of the anomaly carrier phases. Although isotope studies on sequentially digested primitive chondrites can provide some hints (*e.g.*, Burkhardt et al. 2019; Burkhardt et al. 2012; Frossard et al. 2024; Paton et al. 2013; Schiller et al. 2015), identification and detailed nano-characterization of the presolar anomaly carriers will be needed to understand how these grains respond to thermal processing (Nittler et al. 2020). This is not an easy task, since presolar grains are rare. Within chondritic matrices, presolar grains constitute up to a few hundreds of ppm of the total matrix. As a result, to date, only the carriers of $^{54}$Cr anomalies have been successfully identified in the form of presolar spinel (Dauphas et al. 2010; Qin et al. 2010).

Thermal processing of dust has, so far, mainly been considered to drive the nucleosynthetic diversity recorded by meteorites and their components but it may also operate during planetary accretion. Considering this process may allow to resolve the apparent lack of a nucleosynthetic endmember component in the meteoritic record that appears to be required to explain the composition of the Earth in multi-element-multi-isotope plots such as $\varepsilon^{96}$Zr vs. $\varepsilon^{50}$Ti and $\varepsilon^{94}$Mo vs $\varepsilon^{54}$Cr (Fig. 8). In detail, Johansen et al. (2023) presented a model of pebble accretion during which gas-drag assisted accretion of mm-to-cm sized particles onto a growing planet would lead to the inevitable formation of a hot envelope around the accreting planet. Processing of solids that accrete onto a planet with a hot envelope may modify the planet's bulk chemical composition, whereby sublimated solids can be lost to the surrounding protoplanetary disk via recycling flows (*e.g.*, Steinmeyer et al. 2023). Onyett et al. (2023) proposed that the same process may also allow modification of isotope signatures such as those controlled by carriers with extreme nucleosynthetic isotope compositions. Building on the proposal of an inward flux of CI-like dust to accreting planets (Schiller et al. 2020; Schiller et al. 2018), Onyett et al. (2023) proposed that a modest loss, via recycling flows through the envelope, of isotopically "normal" Zr, Nd and Mo relative to refractory and highly *s*-enriched isotope carriers (*e.g.*, SiC grains) could alleviate the need for a hidden reservoir of these isotope signatures in the meteorite record and explain the isotope composition of the Earth. As such, the apparent "missing end-member" needed to explain Earth's isotope signature in Zr, Nd and Mo in inheritance models may instead be indeed "missing" in the form of isotopically normal material that was lost during the Earth's accretion. The interpretation of Onyett et al. (2023) has been challenged, however, as it remains unclear how this process makes the Earth an isotopic endmember for multiple geochemically and cosmochemically distinct elements, and why various differentiated and undifferentiated objects form common NC isotope anomaly trends in multi-elemental isotope space in the first place (Morbidelli et al. 2025). Nevertheless, if interpreted in the framework of additional processing of dust during terrestrial planet formation, the larger implication of the isotope signatures would be that terrestrial planet formation did not follow the classical model of stochastic planetary growth of ever larger collisions, but instead was mainly driven by pebble accretion (Johansen et al. 2021; Johansen et al. 2015).

### 4.3. Dichotomy vs trichotomy

The unique nucleosynthetic anomalies of Fe and Ni in CI chondrites (and Ryugu) are at the origin of an ongoing debate. At the current limits of precision, CI chondrites and Ryugu have identical Fe and Ni isotope anomalies, which are distinct from that of other CC chondrites (Fig. 14): *i.e.*, in any ε-ε plot involving either Fe or Ni isotopes, CIs and Ryugu are positioned well outside the CC field. The chemical affinity between Fe and Ni suggests their anomalies share a common carrier. These anomalies primarily reflect variations in the abundances of nuclides $^{54}$Fe, $^{58}$Ni, and $^{60}$Ni. Relative to other CCs, CIs/Ryugu appear depleted in $^{54}$Fe, plotting close to the BSE in $\mu^{54}$Fe (*see Section 3.3*; also Fig. 8), but enriched in $^{58}$Ni and $^{60}$Ni. This departure from a coupled excess in the neutron-poor nuclides $^{54}$Fe and $^{58}$Ni, which defines the CC offset from NCs, suggests the parent bodies of CIs/Ryugu sampled a distinct carrier during their formation, deriving from a region and/or time in the protosolar disk wherein (i) the carrier was uniquely present (*i.e.*, a trichotomy in disk isotopic reservoirs; Hopp et al. 2022a), and/or (ii) astrophysical conditions (*i.e.*, the mode of planetesimal formation; Spitzer et al. 2024) were conducive to its incorporation.

An isotopic trichotomy was initially proposed on the basis of Fe anomalies in Ryugu (Dauphas et al. 2024; Hopp et al. 2022a). In particular, the composition of CIs/Ryugu was posited to reflect that of a rarely sampled reservoir lying beyond the CC formation zone, around the birthplaces of Uranus and Neptune. Some subsequent works (Marrocchi et al. 2023; Yap and Tissot 2023) suggested the appearance of a trichotomy resulted from a lack of Fe isotope data, predicting that CC achondrites would bridge the gap between CIs/Ryugu and the other CCs, thereby reaffirming CIs as an endmember in the CC intra-reservoir trend. Analyses of Fe anomalies in those achondrites have since falsified this prediction (Rego et al. 2024), establishing the distinct genetic heritage of CIs/Ryugu as a true feature. Subsequent Ni anomalies measured in Ryugu, CIs, and CCs corroborated the special status of CIs/Ryugu (Spitzer et al. 2024), and inspired the suggestion that micron-sized Fe-Ni grains (distinct from the bulk inventory of chondritic iron such as that hosted in chondrules) constitute the carrier of Fe and Ni anomalies. As planetesimal formation by hydrodynamic instabilities (*e.g.*, the streaming instability) tend to feature the largest (typically mm-cm in size; Yap and Batygin 2024), and thus the most aerodynamically coupled, dust grains in the disk locality (*e.g.*, Squire and Hopkins 2018; Youdin and Goodman 2005), these micron-sized Fe-Ni grains are envisioned to have contributed minimally to the building blocks of most planetesimals, such as the parent bodies of the CC chondrites. Towards the end of the protosolar disk lifetime (~ 4Myr from CAIs; Borlina et al. 2022; Wang et al. 2017), photoevaporation facilitates gas dissipation and with it, a rise in metallicity across the disk. This is thought to trigger a final wave of planetesimal formation, not by said hydrodynamic instabilities, but through direct gravitational collapse (*e.g.*, Goldreich and Ward 1973). Such collapse does not involve grain size-sorting, thus incorporating the anomalous micron-sized Fe-Ni grains in abundance. CI chondrites and Ryugu are suggested to originate from this last generation of planetesimals.

Physical interpretations for the observed systematics, while the ultimate goal, remain largely unsubstantiated. Indeed, the suggested anomalous Fe-Ni grains have yet to be identified (*i.e.*, shown to exist), isolated, and studied. At present, all that can be inferred with certainty is that CIs and Ryugu accreted in a different place and/or time than the other CCs.

## 5. Evidence for disk transport and mixing: Insights from isotope anomalies in nebular dust components

Observations of young stellar objects reveal that the amount of crystalline to amorphous silicate dust grains in outer disk regions is higher than in the ISM. This is interpreted as evidence for the outward transport of crystalline materials formed by condensation or thermal annealing of amorphous dust close to the protostar (Watson et al. 2009). Similar transport and mixing of primitive and processed dust also happened in our SS, and the investigation of cometary materials and chondritic meteorites provides clues for reconstructing these processes in some detail.

### *5.1 Cometary materials*

Comets and chondrites can be understood as non-equilibrium "cosmic sediments" sampling different depositional environments within the evolving disk. Like sediments on Earth, their composition, makeup, and texture provide constraints on the genetic heritage of their constituents, as well as on transport, sorting, and alteration processes on the way to – and within – their respective accretionary environments. In this regard, comets are cosmic sediments from the cold distal regions of the outer SS. Yet, the laboratory analysis of material collected from the coma of comet Wild-2 in the Stardust mission revealed a surprisingly high abundance of high-T components such as refractory inclusions and olivine. These observations have been interpreted as evidence for the transport of early-formed crystalline solids from the inner disk to the accretion region of comets (Brownlee 2014). Furthermore, isotopic analyses of the comet dust grains showed that the transported materials stem from isotopically distinct source reservoirs. In particular, the O-isotopic compositions of refractory crystalline solids imply that the outward-transported cometary dust formed at different radii in the disk, $^{16}$O-enriched grains close to the protosun, and $^{16}$O-depleted grains further out, possibly near the accretion region of chondrites (McKeegan et al. 2006). Together, these results led to the realization that the outward transport of dust is a natural outcome of early disk evolution (see *Section 4.1*), and that even highly primitive SS objects like comets are mixtures of processed and unprocessed disk materials of diverse heritage.

### *5.2 Chondrite components*

Stardust samples aside, the main source of information for investigating transport, mixing, and processing of dust in the solar accretion disk are chondrite meteorites. They come in much larger sample sizes compared to the few returned cometary grains, and, importantly, in a range of different flavors, each thought to represent a specific snapshot of the materials present in the protosolar disk at the time and place of their parent bodies' accretion. The NC chondrites are thought to have formed ~2 Myr after CAIs in the inner regions of the SS, close to the formation region of the terrestrial planets, while the CC chondrites formed ~2-4 Myr after CAIs in further out regions, with proposals ranging from beyond Jupiter's orbit to beyond Neptune's orbit (Fig. 11). The main constituents of chondrites are the eponymous chondrules (sub-mm to mm-sized once-molten, fast-cooled silicate spherules formed by unclear processes in the presence of nebular gas; Russell et al. 2018), opaque phases (*e.g.*, metal and sulfides), refractory inclusions (CAIs and AOAs), and a fine-grained matrix consisting of crystalline and amorphous solids, fragments of other chondrite components, organics, and presolar grains. Variations in the relative abundances and physico-chemical characteristics (*e.g.*, size, mineralogy) of the main chondrite components form the basis for the phenomenological classification of chondrites into different groups and types, and testifies to the distinct/changing accretion environments in the disk (*e.g.*, Scott and Krot 2014). Understanding the origin and relation of the different components within and among chondrites is key for understanding the mechanisms and extent of disk transport and processing. To date, much of the cosmochemical debate related to this point has centered around the question of whether the main components of a given chondrite formed mostly by fractionation from an isolated local disk volume, or whether they were transported over wide distances from separate disk regions to mix into the accreting parent body. As conservative source tracers, isotope anomalies in meteorite components provide constraints for teasing apart these scenarios and better understand transport, mixing, and accretion processes in the disk.

*Refractory inclusions:* Clear evidence for the transport of material from different disk source regions to the accretion region of chondrites comes from the isotopic analysis of refractory inclusions. For many elements, CAIs and AOAs have relatively uniform, but highly anomalous isotopic compositions compared to the composition of their host chondrites (*e.g.*, Jansen et al. 2024; Torrano et al. 2024). This necessitates a different source reservoir of refractory inclusions compared to other chondrite components, and – like for the CAIs found in the Stardust samples – implies an outward transport of early-formed phases. Moreover, within the different classes and types of chondrites, the abundance of refractory inclusions is highly variable. While CAIs are very rare in NC chondrites (<0.01 area %, Dunham et al. 2023), they constitute up to ~5 area % of CC chondrites (Hezel et al. 2008). This heterogeneity implies variability in the distribution of refractory inclusions in space and time, which has been linked to the formation of chondrite parent bodies within pressure bumps in the disk (Desch et al. 2018; Hellmann et al. 2023). More specifically, these models suggest that the CC chondrites formed through the pile-up of inward-drifting particles at different times and/or different distances in pressure bumps outside of proto-Jupiter. These particles include CAIs transported to the outer disk (see *Section 4.1*). In this scenario, the near absence of CAIs in NC chondrites is the result of both trapping of CAIs at pressure bumps, and radial drift of inner SS CAIs into the protosun early on.

An important implication of the variable abundance of refractory inclusions in chondrites is that it exerts a significant control on the bulk isotope anomalies and chemical compositions of CC chondrites. This is evidenced by trends between the abundances of CAIs, the abundances of refractory elements like Ca, Ti, Sr, Zr, or Nd, and the bulk CC chondrite isotope anomalies (*e.g.*, Burkhardt et al. 2019; Yap and Tissot 2023). Moreover, these trends pass through the compositions of CC chondrules and end at the composition of NC chondrules (Fig. 2 in Gerber et al. 2017). This implies that CAIs and AOAs are not only present as entities in carbonaceous chondrites, but also have been part of the precursor materials of CC chondrules.

*Chondrules:* As the dominating component in most chondrites, chondrules likely played a critical role in mediating the accretion of dust into planetesimals (*e.g.*, Budde et al. 2016; Scott and Krot 2014). Determining the origin of chondrules within a given chondrite, as well as among different chondrite groups and types is thus of prime interest for investigating dust transport in the disk, and ultimately provides clues about the nature and site of chondrule formation and planetesimal accretion. Isotopic variability among and within chondrules has been observed for a number of elements with different cosmochemical properties, including O, Ti, and Cr (*e.g.*, Marrocchi et al. 2024b; Olsen et al. 2016; Schneider et al. 2020; Van Kooten et al. 2021; Van Kooten et al. 2020; Van Kooten et al. 2016; Williams et al. 2020).

Oxygen is nominally a volatile element, but in oxides and silicates also part of refractory and main component minerals of nebular dust. Given these attributes, the O isotopic composition of individual chondrules is the result of a combination of precursor heterogeneity, open system gas-melt interaction during chondrule formation, and mineral-water/ice interaction during aqueous alteration. Individual chondrules of a given chondrite type show some spread in O isotopic composition, and there is overlap between chondrules from different chondrite groups and types (Clayton et al. 1991). This means that analyzing the O isotopic composition of a single chondrule does not allow one to assign it to a certain chondrite type with high confidence. This begs the question as to whether individual chondrules in a given chondrite (i) formed locally from isotopically variable precursor mixtures, or rather (ii) originated from isotopically distinct regions in the disk and were transported over large distances to the chondrite accretion region.

Support for the former interpretation comes from the observation of relict grains in chondrules. These grains are found mostly in CC chondrules and are characterized by $^{16}$O-rich compositions relative to the composition of the host chondrule (Ushikubo et al. 2012). The identification of relict grains imply that CAIs and AOA-like materials were part of the precursor mix of the chondrules, and hence, that isotopic heterogeneity among individual chondrules is inherited from heterogeneities at the chondrule precursor scale, and not necessarily a result of disk-wide chondrule transport (Marrocchi et al. 2018). In contrast, the recent finding of an O isotopic heterogeneity in ordinary chondrite chondrules that scales with chondrule-size may be interpreted as evidence for the transport of some small $^{16}$O-rich chondrules from the CC to the NC reservoir, and hence a leaky NC/CC barrier (Marrocchi et al. 2024b). Isotope data for other elements will be needed to confirm an outer SS origin for these chondrules.

Chromium is a main component element, such that unlike for O, the Cr concentration and isotopic composition of chondrules is not sensitive to gas-dust interaction, but thought to trace the bulk dust component of the disk. As such, the observation of Cr isotopic variability among individual CC chondrules spanning the range of bulk NC to CC bodies was interpreted as indicating large-scale transport of chondrules from different disk regions to the accretion region of CC chondrites (*e.g.*, Olsen et al. 2016). Likewise, the observation of pristine CC chondrules with NC-like ε$^{54}$Cr isotope signatures in their cores, and CI-like compositions towards their rim were interpreted as signatures of massive outward transport of NC chondrules to the outer SS, potentially coupled with inward flux of CI-like material to the NC region (Van Kooten et al. 2021). Elevated ε$^{54}$Cr in two chondrules from an enstatite chondrite support transport from the outer to the inner disk (Zhu et al. 2020). Moreover, the ε$^{54}$Cr data of the individually analyzed chondrules form a continuum filling the isotopic "gap" between NC and CC bulk bodies. Assuming that the ε$^{54}$Cr composition of the chondrules represents solely a source signature would imply that there are chondrules sampling disk regions between the NC and CC reservoirs, where no bulk planetary body formed (or at least for which no bulk samples are available in the world's collections). Alternatively, the Cr isotope anomalies could also indicate isotopic heterogeneity at the chondrule precursor scale, *i.e.*, heterogeneous mixing of NC and IC materials during chondrule formation in localized disk regions.

Titanium is a highly sensitive tracer of isotopic heterogeneity at the chondrule precursor scale with respect to the presence of CAIs. This is because CAIs have large Ti isotope anomalies and are strongly enriched in Ti compared to chondrules. Furthermore, in contrast to Cr (Grossman and

Brearley 2005; Van Kooten et al. 2021), the Ti isotopic composition of individual chondrules is not subject to later modification by aqueous alteration on the parent body. Hence, the Ti isotopic variability observed in CC chondrules was readily assigned to variable admixing of CAIs to a NC-chondrule-like starting material at the individual chondrule level (Gerber et al. 2017; Niemeyer 1988b). This implies that NC materials and CAIs have been present in the formation region of CC chondrules, in line with the proposition that the CC reservoir represents a mixture of NC and IC materials.

Such a mixing scenario is also preferred in combined O, Ti, and Cr isotope studies of NC and CC chondrules (Fig. 12), which reveal a restricted range of isotopic compositions for NC chondrules, and a wide spread in CC chondrules, from the NC field towards the composition of CAIs and AOAs (Schneider et al. 2020; Williams et al. 2020). Furthermore, these data indicate that although isotopically NC-like material is present in NC chondrules and CC chondrules, the formation region of the NC and CC chondrules was likely distinct. This is because refractory-enriched chondrules from NC and CC chondrites have a contrasting isotopic composition, indicating a distinct origin of the refractory materials in the NC and CC reservoir (Ebert et al. 2018). This has been interpreted as supporting the existence of a strong barrier between the outer and inner SS, and a limited inward drift from the CC to the NC reservoir. However, more multi-elemental isotope anomaly data of a more diverse set of chondrules from different NC and CC chondrites are needed to address the extent of chondrule transport between the disk reservoirs, and hence the nature and permeability of the barrier more confidently.

### *5.3 Mixing and transport: the current picture painted by nebular dust components*

Although knowledge gaps remain, the isotope anomaly data from cometary and chondritic materials reveals that outer SS bodies are mixtures of isotopically diverse materials at all scales. Carbonaceous chondrites are mixtures of multiple components (*e.g.*, CAIs, chondrules, matrix), some of which are themselves mixtures of materials of distinct nucleosynthetic heritage (*e.g.*, Desch et al. 2018; Hellmann et al. 2023). This is particularly evident from the isotopic composition of CC chondrules, which are mixtures of NC materials and variable amounts of refractory inclusion and CI-like dust (Fig. 12). CI-chondrites (and comets) are, in turn, themselves mixtures of low and high-temperature materials of also distinct isotopic heritage (*e.g.*, Barrat et al. 2016; Dauphas and Pourmand 2015; Frank et al. 2023; Kawasaki et al. 2022). Taken at face value, these observations call into question the well-accepted idea that CI chondrites are faithful proxies of the chemical and isotopic composition of primitive SS dust (see *Section 4.1 ii*). At a higher level, they also highlight how all SS materials, even those traditionally thought of as highly primitive (comets, CI chondrites), are intimately linked and testify to the details of disk dynamics. The continued exploration of nucleosynthetic anomalies in meteorite components (which is currently in its infancy, Fig. 5) promises to bring more granularity, and hopefully clarity, to the emerging picture of disk wide transport and mixing of dust components.

### **6) Limitations, open questions, and outlook**

As discussed above, the careful investigation of nucleosynthetic anomalies in CAIs, bulk meteorites, and more recently AOAs, chondrules, and matrices have brought the most important constraints to-date on our understanding of the architecture and evolution of our early SS. This community effort over decades has now yielded a rich dataset, spanning more than 20 elements, including all geo- and cosmochemical categories, and revealing systematics behaviors of isotopic anomalies across them. At the same time, these momentous advances have also unveiled major questions.

First and foremost is the question of the origin of the isotopic heterogeneity observed in SS materials. The debate currently focuses on the *inheritance* and *unmixing* models (Fig. 11 & 13), which both have their respective strengths and weaknesses. The major strength of the inheritance models is that they explain with a minimum number of free parameters (3 in total: 2 for the NC reservoir and one for the IC reservoir) the global systematics seen across all elements investigated to date. Qualitatively, this is a remarkable feat, and initial quantitative testing of such models (Yap and Tissot 2023) showed that the elemental and isotopic composition of CC chondrites can be consistently reproduced within ~20 % for the four elements investigated (Ca, Ti, Cr, Fe). Another strength of the inheritance models is that they are physically grounded into the astronomical observations of filaments, streamers and other (sub)structures in planet-forming disks (see *Section 2*). An obvious limitation of the inheritance models is that, while the reservoirs that they invoke are physically motivated – for instance, the IC reservoir's isotopic composition is that of CAIs and AOAs while its chemical composition is chondritic – they are nonetheless hypothetical: *i.e.*, there are no known bulk meteorite samples of the actual reservoirs. This is best encapsulated by the denomination of the "missing end-member" (Burkhardt et al. 2021; Burkhardt 2021), which describes the reservoir needed to explain the terrestrial composition, and without which the Earth ends up being an end-member itself in many isotope-isotope spaces (*e.g.*, Fig. 8b-e). This also begs the question of whether such samples actually exist, *i.e.*, for how long the primary reservoirs persisted before their original signatures were diluted to the compositions recorded in planetary materials sampled today. Indeed, recent isotopic analyses of outer disk materials suggest that the comet-forming region has not been dominated by an IC component at the time of dust accretion, but that a streamer likely contributed primitive molecular cloud material to this region after the initial disk formation and before parent body assembly (Van Kooten et al. 2024). In the inheritance models, physico-chemical processing of SS materials is secondary to mixing and transport within the disk.

In contrast, the unmixing models do not invoke hypothetical reservoirs but hypothesize that CI chondrites, which are a known family of meteorites, represent the unprocessed primitive materials from which the SS formed. Furthermore, they build on the idea that physico-chemical gradients in the disk fundamentally impact the elemental and isotopic make-up of the evolving protoplanetary disk. Clearly, a strength of these models is that they build on the processing of a known starting material to explain the observed heterogeneity in SS materials. At the same time, and in contrast to the simplicity of the mixing processes invoked in the inheritance models, unmixing models must rely on a large number of free parameters to explain the isotope anomalies systematics in bulk meteorites and across elements with different geo- and cosmo-chemical behaviors. This is a major limitation of these models as the carriers of different types of anomalies, which are today correlated in bulk meteorites, are likely to respond differently to thermal processing. Furthermore, and as discussed in *Section 5*, new evidence is showing that CI chondrites themselves are mixtures of low- and high-T materials with distinct isotopic heritage. As such, they are unlikely to be good proxies of the unprocessed primitive SS composition.

At the heart of the debate between the inheritance and unmixing models loom some key unknowns. The first is our nearly total lack of knowledge on the nature and physico-chemical properties of the carriers of nucleosynthetic anomalies. Since the processing of these carriers is key to the actual mechanisms invoked by unmixing models, testing in detail these models remains extremely arduous, if not impossible. Efforts devoted to the identification and characterization of nucleosynthetic anomaly carriers (*e.g.*, Dauphas et al. 2010; Qin et al. 2010) will be critical to future progress. A second issue pertains to identifying whether the trends seen in isotope-isotope space testify to a temporal and/or spatial variability of the protoplanetary disk. Current chronological constraints mainly rely on accretion ages of iron meteorites, and seem to indicate that both NC and CC irons formed coevally, pointing to a spatial separation of NC and CC materials. But accretion ages come with built-in assumptions about the size and composition of their respective parent bodies, which might not necessarily be correct.

Clearly, the inheritance and unmixing models are end-members scenarios on what must in reality represent a spectrum of possibilities, with both processes contributing to different extents to shaping the evolution of the SS. Given the current state of our knowledge on isotope anomalies and their carriers, Occam's razor points to the inheritance models as being most appealing since (i) they can explain the current data with the least number of free parameters (only 3 components needed), and (ii) astronomical observation have revealed a ubiquity of heterogeneous (sub)structures in star- and planet-forming regions, which provide a viable mechanism for delivery of compositionally heterogeneous material to the protosolar disk throughout its formation and until the latest stages of infall.

It is important to recognize that to elucidate the origin of the isotopic heterogeneity of SS materials, two complementary avenues can be pursued. The first, and most common one in the literature so far, is to focus on the physical interpretation for the observed isotopic systematics. That is, to devise models for where, why, and how heterogeneity in the early SS arose and was maintained, and how that heterogeneity was sampled by parent bodies formed across the disk across its lifetime. Another, less popular but arguably more neutral, path forward is to identify the number and nature of endmembers necessary to create the emergent systematics seen in multi-element isotope space. These endmembers would correspond to nebular dust components that mixed (*Section 4.1*) and/or unmixed (*Section 4.2*) to yield the isotopic diversity in SS materials (Yap and Tissot 2023). Although one step removed from the physical, the latter avenue serves as a logical prerequisite to the latter.

At the intersection of meteoritics, astrophysics and astronomy, another important open question is that of the efficacy of disk transport, in particular in the earliest stages of disk formation and evolution. The presence of refractory crystalline solids in cometary samples and carbonaceous chondrites is in line with observations of high crystalline to amorphous dust ratios in disks of YSOs, and testifies to the outward transport of materials forming in high-T settings close to the young star. However, the nature and efficiency of this transport process remains debated, with proposed mechanisms ranging from viscous disk spreading (Ciesla 2007), large scale stochastic diffusion (Morfill and Voelk 1984), and ballistic orbits above the disk plane through X-winds (Shu et al., 2001). Further study of isotope anomalies in meteoritic components will provide additional insights to help differentiate these possibilities. For instance, the detection of excess nucleogenic isotope variations in CAIs or chondrules would be a fingerprint of irradiation during transport above the disk midplane, while the absence of any such excess would suggest embedded transport, and hence favor viscous spreading or stochastic diffusion. Exploratory work in this direction has thus far returned no positive detection, in support of the latter scenarios (Ghaznavi et al. 2025; Roth et al. 2016). Likewise, the detection of materials with an isotopic composition similar to that of CAIs and AOAs, but a less-refractory chemical composition would provide additional support for the viscous disk spreading (and inheritance) model, while the absence of this hypothesized IC material would be in line with the X-Wind model. A cometary sample large enough to obtain high-precision isotope anomaly data for a number of elements, in addition to the outer disk xenoliths already observed in various chondrite groups (Goodrich et al. 2021; Kebukawa et al. 2023; Van Kooten et al. 2024; Van Kooten et al. 2017b; Van Kooten et al. 2017a), would help to resolve these debates (Bockelée-Morvan et al. 2022).

Another open issue with profound impact on our understanding of disk transport and mixing is the question whether or not the components found in chondritic meteorites represent the materials also forming the first generation of planetesimals. More specifically, it remains unclear whether achondrite parent bodies initially contained similar mixtures of refractory inclusions, chondrules, and matrix as found in chondrites. The observation of relict chondrules in primitive achondrites (Schrader et al. 2017) and similar isotope anomaly ranges in chondrites and achondrites seems to support this notion (Fig. 8), however, based on trace element patterns a recent work also argued that some iron meteorite parent bodies might have formed with up to 25 wt% CAIs (Zhang et al. 2024). If correct, this would require efficient trapping, sorting, and enrichment mechanisms in the disk, and would imply the existence of CAIs with isotopic compositions different from the ones found in carbonaceous chondrites. Such CAIs with a more NC-like composition might have existed (Brennecka et al. 2020; Ebert et al. 2018), but more work is needed to pinpoint them as refractory products of the NC reservoir. CAIs aside, the most abundant component in chondrites are chondrules, and thus the question of whether chondrites and achondrites share the same starting material is directly related to the unresolved debate about the nature and timing of the chondrule forming process(es), and hence the importance of chondrules in protoplanetary disks (Marrocchi et al. 2024a; Piralla et al. 2023; Russell et al. 2018). Clearly, future work on isotope anomalies in chondrules of different chondrite types, in particular with respect to apparent trends with chondrule size (Gerber et al. 2017; Marrocchi et al. 2024b; Marrocchi et al. 2022), as well as studies of matrix and metal phases of different chondrites will provide novel perspectives for linking SS dust processing and transport to astronomical observations and astrophysical concepts of disk evolution.

Our capacity to interpret the message from meteorites also depends on future breakthroughs in astronomical observations and theory, which inform our understanding of the early SS by using protoplanetary systems as analogues to the early protosolar disk. With a large sample of sources, current observations provide snapshots of disks (i) at different evolutionary stages, (ii) within a variety of star-forming environments, and (iii) undergoing stochastic events, such as accretion outbursts or infalling streamers. It remains uncertain, however, which conclusions drawn from studying these other protoplanetary systems are directly applicable to the case of the early SS. Timelines of events from astrophysical observations are difficult to pin down; the uncertainties on protostellar ages are typically on the order of several million years. Streamers themselves have only recently become an area of study, with a growing number of detections in molecular line observations of protoplanetary sources. With no unbiased samples from dedicated searches to date, streamer detections have mostly happened accidentally, in existing ALMA data, which are themselves heavily focused on low-mass star-forming regions. This should hopefully change in the near future as strategies for dedicated searches for new observational campaigns are being developed and candidate sources are identified in archival data (Gupta et al. 2023). Such future investigations will be necessary to constrain the streamer occurrence rates in different star-forming environments and tackle the currently open question of the role of streamers in planet formation and their impact across the lifetime of the protoplanetary disk.

Finally, advances in the resolution of astronomical observations will be key to better understanding the process of planetary formation. To date, only the reservoirs of planetary ingredients, both solids and volatiles, have been mapped with high resolution observations (~10 AU) at (sub)millimeter wavelengths for a number of nearby systems. In contrast, the signposts of planet formation (*e.g.*, dust substructures, embedded protoplanets) identified so far have been limited to those at much larger separations (10s of AU) than that of the SS planets with inferred masses of protoplanet candidates typically on the order of several Jupiter masses. Even with current facilities, knowledge of the inner 10-30 AU of the disk midplane is limited; at millimeter wavelengths, the midplane dust emission is often optically thick and not spatially resolved, preventing an accurate accounting or characterization within much of the region of interest for comparison to the protosolar nebula. Future facilities, however, such as the proposed next generation Very Large Array (ngVLA) are slated to be capable of 5 milliarcsecond angular resolution (sub-AU resolution for sources within 140 pc) and higher sensitivity at longer wavelengths than ALMA where dust emission is less likely to be optically thick (van der Marel et al. 2018). At these capabilities, the ngVLA would be able to resolve the substructures induced by > 5-10 Earth mass protoplanets at separations of ~ 5 AU (Ricci et al. 2018) – paving the way for unprecedented observations such as, for instance, the detection of the signposts of a proto-Jupiter captured in mid-formation.

## Declarations

**Competing Interests** The authors declare that they have no conflict of interest.

**Acknowledgements** We thank Steve Desch and Yves Marrocchi for constructive reviews that helped improve the manuscript, and Alessandro Morbidelli for prompt and careful editorial handling. We are grateful to workshop organizers Herbert Palme, Jutta Zipfel, Alessandro Morbidelli, Klaus Mezger and Dominik Hezel for their invitation and leadership, and to the International Space Science Institute (ISSI) for supporting and hosting this event. This work was further supported by a Packard Fellowship (to FLHT).

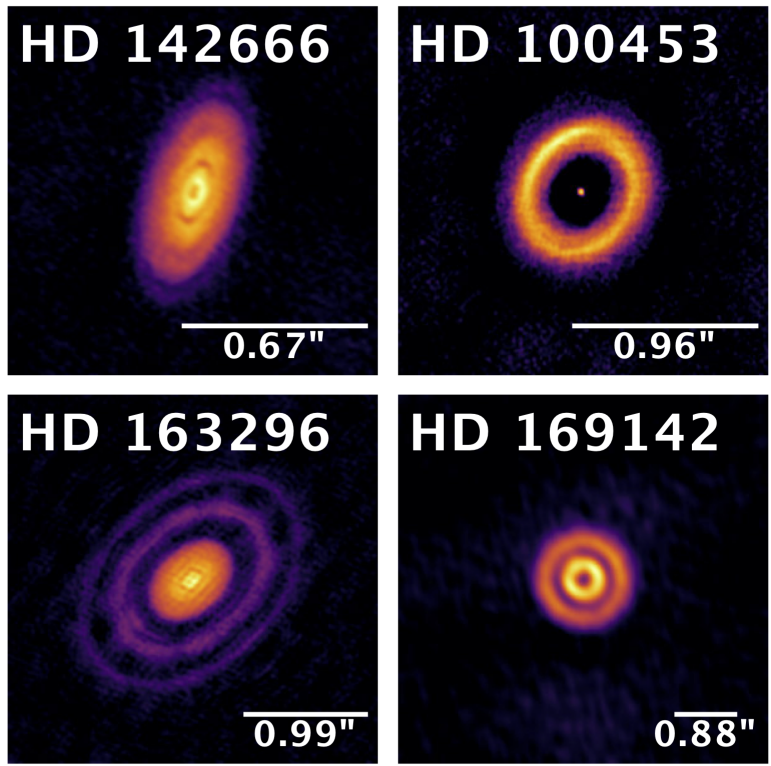

**Fig. 1. A selection of ALMA images (Band 6 and 7 continuum) illustrating the diversity of sizes and structures of circumstellar disks.** Scale bar is 100 au (value denotes the angular scale in arcseconds). Images from Stapper et al. (2022).

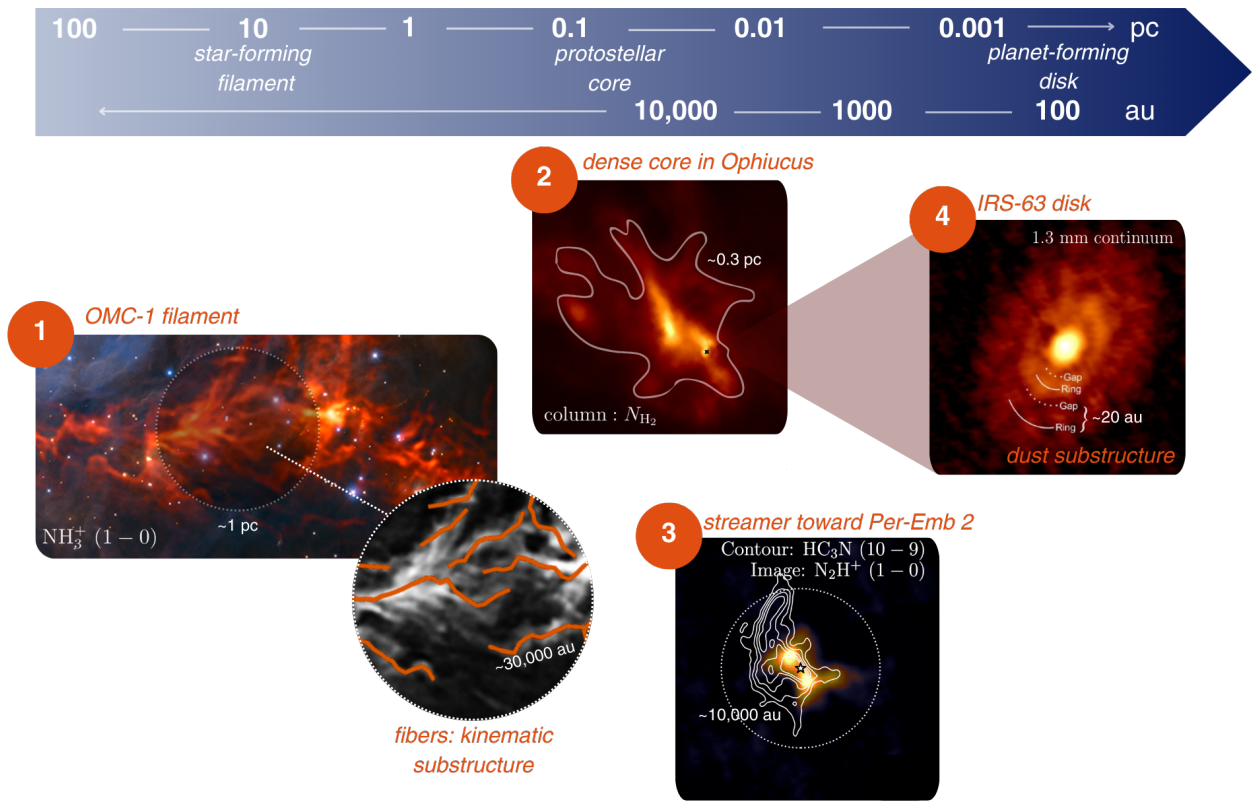

**Fig. 2. Filamentary structures and anisotropy are observed across the dynamic range of star formation.** Starting at the star-forming scale, (1) shows the OMC-1 star-forming filament observed by ALMA+IRAM 30 m in the form of 'fibers' identified as velocity coherent on 30,000 au scales (Hacar et al., 2018). Such filamentary behavior and anisotropy is associated with the presence of (2) dense protostellar cores, as in the heterogeneous column density of L1709, inferred from Herschel observations (Arzoumanian et al., 2019) and fit with ballistic free-fall trajectories as in the case of (3) the 10,500 au 'streamer' structure around the protostar Per-Emb 2 seen in $HC_3N$ by ALMA (Pineda et al., 2020). The structured infalling envelopes of cores like L1709 can be associated with the presence of substructure in the (4) thermal dust continuum (1.3mm) emission from ALMA of its young planet-forming disk (< 500 kyr old) IRS-63 (Segura-Cox et al., 2020).

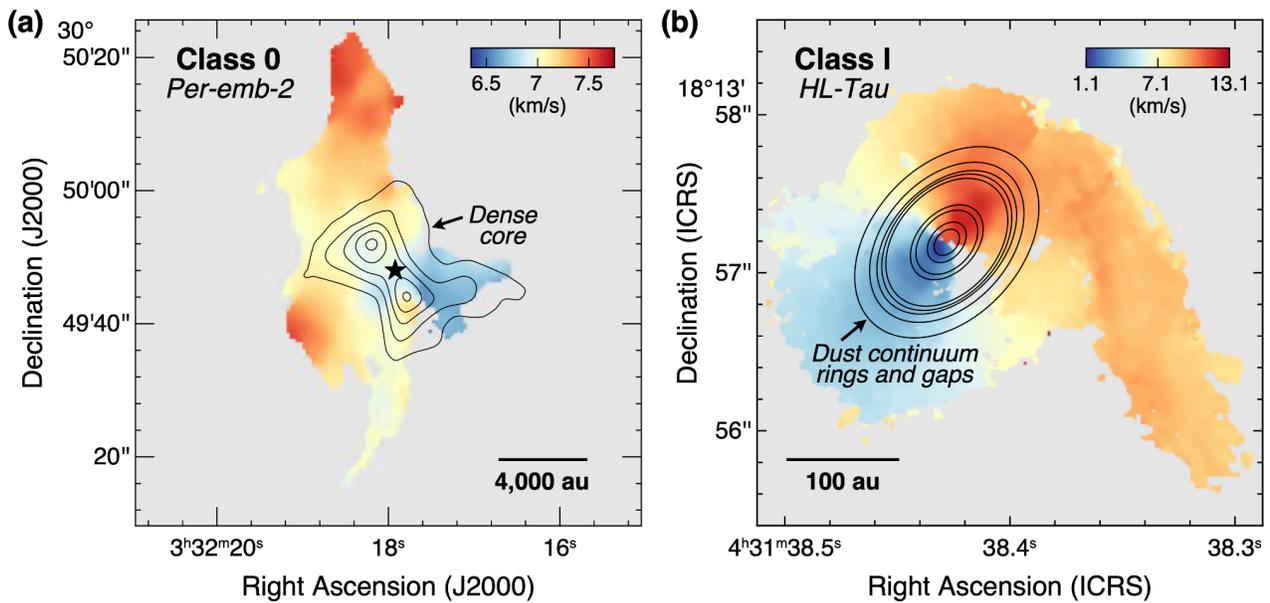

**Fig. 3. Examples of streamers at two different scales corresponding to two classes of young stellar objects (YSO) (modified from Pineda et al. 2022).** (a) Large scale streamer and dense molecular cloud core around the Per-emb-2 Class 0 YSO. The dense core was traced using $N_2H^+$ (1-0), and the streamer centroid velocity (color bar) using $HC_3N$ (10-9) from NOEMA observations. (b) Streamer funneling material onto the disk of the Class 1 YSO HL-Tau. The centroid velocity is derived from $HCO^+$ (3-2), and the disk structure (rings and gaps) is as identified by ALMA Partnership et al. (2015). Streamers could facilitate the delivery of compositionally variable materials into protoplanetary disks.

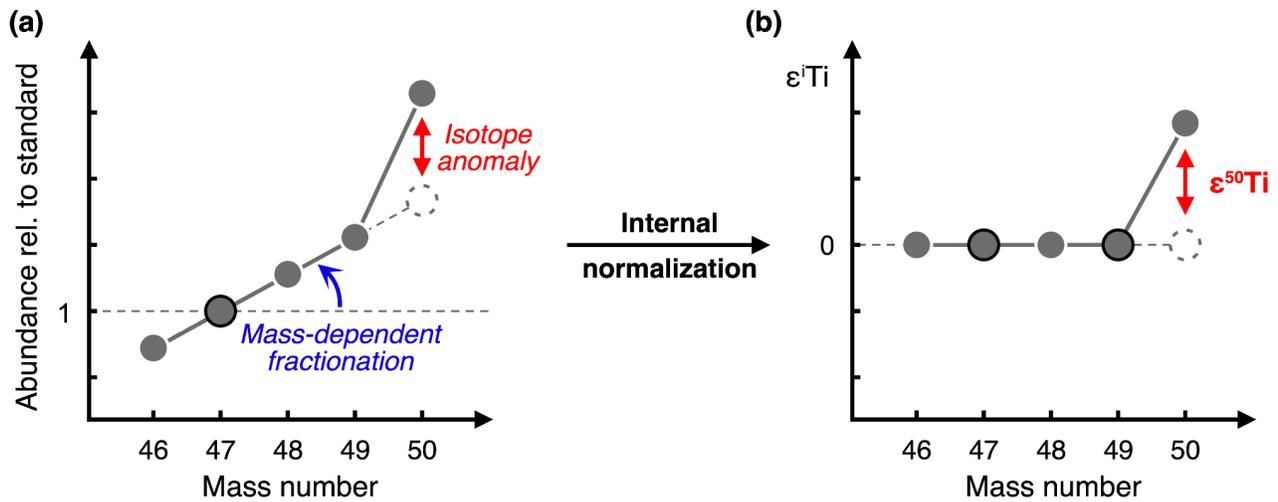

**Fig. 4. Illustration of the difference between mass-dependent vs mass-independent isotope effects.** In panel (a) the y-axis represents the abundance of the various titanium isotopes in a hypothetical sample (normalized to the abundance of $^{47}$Ti) divided by the same ratios in a terrestrial standard. Values on the y-axis are thus the ratio of the $^x$Ti/$^{47}$Ti in the sample divided by the same ratio in the standard, with x being any given isotope. The linear array defined by the isotopic abundances of most Ti isotopes (46 to 49) indicates that the sample experienced some degree of mass-dependent fractionation compared to the standard. The departure from this array of the abundance of $^{50}$Ti (which is highly exaggerated for clarity), represents the influence of a mass-independent process. Panel (b) shows how, after internal normalization (see for instance Russell et al. 1978 and Marechal et al. 1999), the impact of mass-dependent effects has been removed. The y-axis in panel (b) represents the deviations of the internally normalized composition of the sample relative to that of the standard, in epsilon notation (or parts per ten thousands).

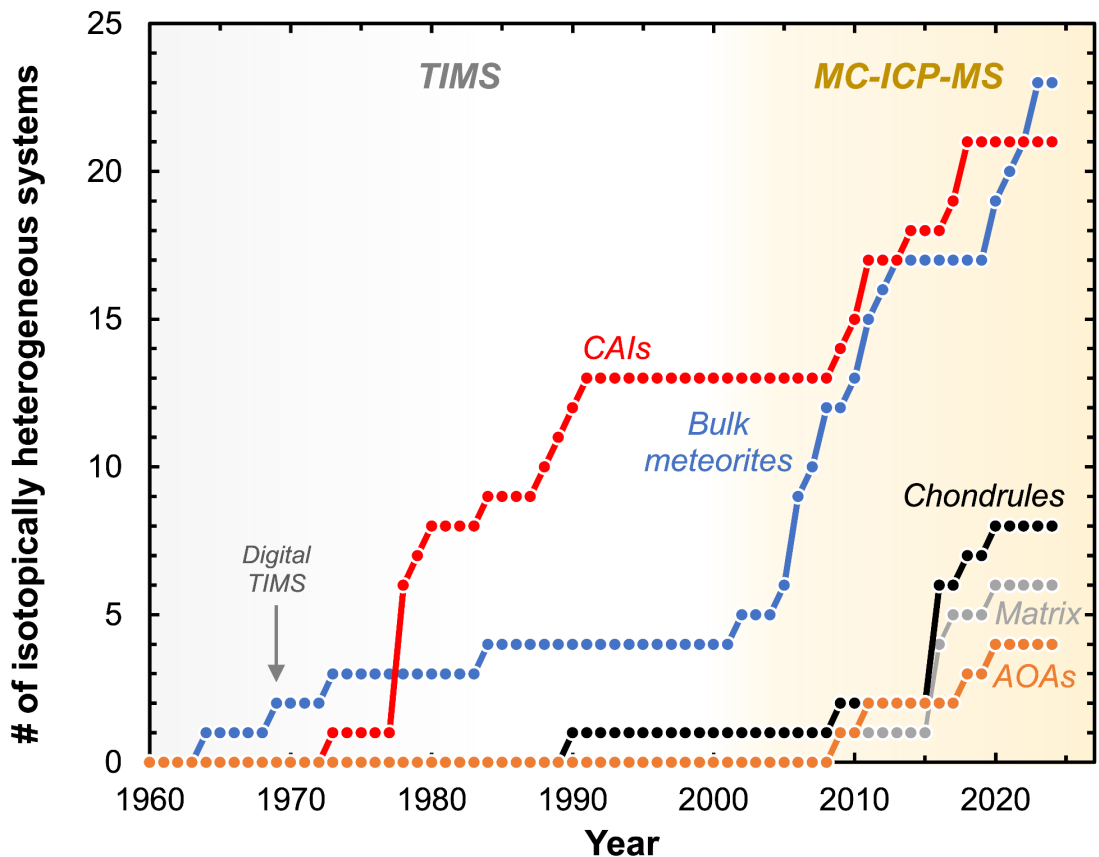

**Fig. 5. Temporal evolution of the number of systems in which isotopic heterogeneity has been documented for bulk meteorites (blue), and their components: CAIs (red), AOAs (orange), chondrules (black), and matrix (grey).** Gas-source mass spectrometry is key for work on noble gases, O, and S, and early work on Si. Following the advent of digital TIMS instruments, isotopic anomalies were rapidly discovered for numerous refractory elements in CAIs (~1980s). Two-decades later, the introduction of MC-ICPMS instruments led to a similar rise in the discovery of isotope anomalies in bulk meteorites. This temporal delay reflects the much larger magnitude of isotope anomalies in CAIs (typically on the order of a few to 10 ε in normal CAIs, and up to 100s of ε in FUN CAIs) compared to bulk meteorites and planets (typically on the order of 0.1 to a couple of ε). Investigation of anomalies in chondrules, matrix, and AOAs has begun only recently. See Supplementary Materials for data source.

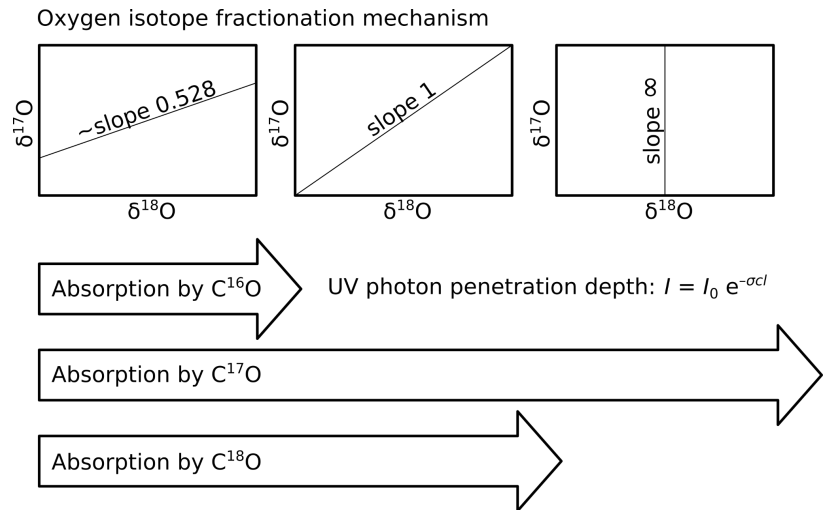

**Fig. 6. Sketch illustrating the effect of self-shielding of CO (modified after Thiemens and Lin 2021).** The penetration depth of the isotopologue-specific photons is a function of the concentration of the isotopologues (Lambert-Beer law). In the region of dissociation of all three CO isotopologues, no or only mass-dependent fractionation occurs. At greater penetration depth, only $C^{17}O$ and $C^{18}O$ are dissociated and the fractionation has a slope of one. At even greater depths, where only $C^{17}O$ is absorbed, fractionation would follow a vertical slope.

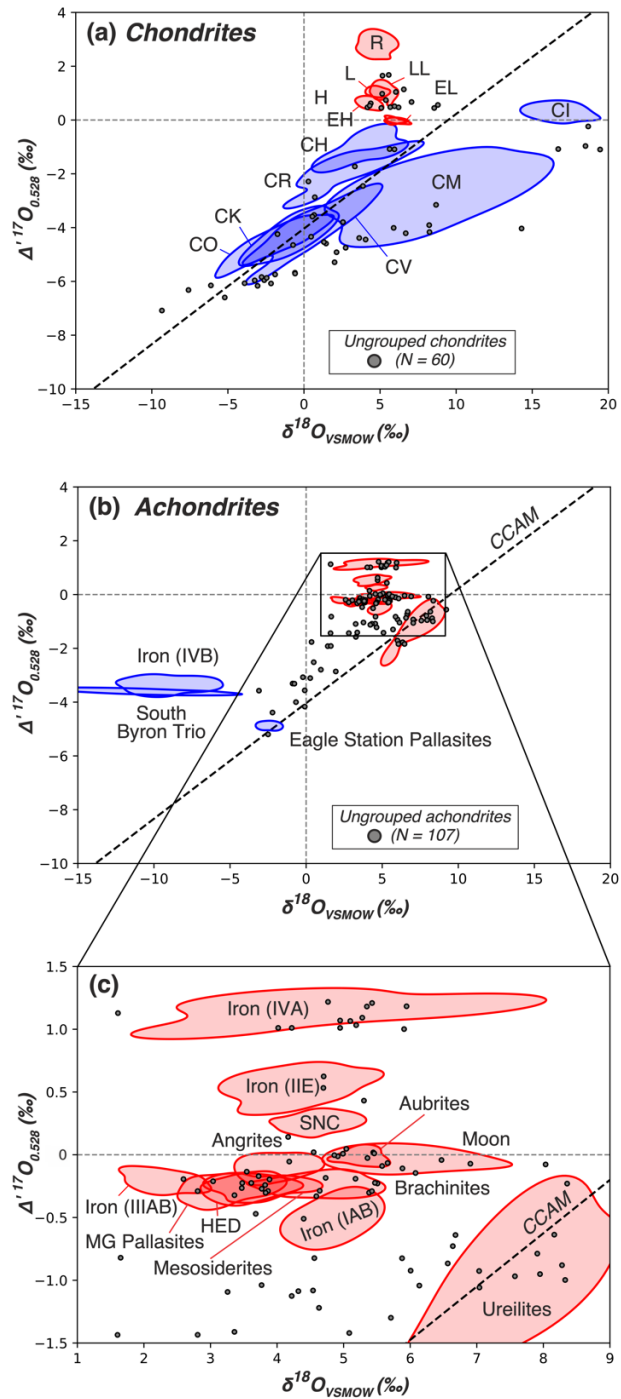

**Fig. 7. Diagrams of mass-independent ($\Delta`^{17}O$) against mass-dependent ($\delta^{18}O$) oxygen isotope signatures of bulk meteorites.** Compilation of chondrites (a) and achondrites (b, c) bulk rock O isotope data extracted from the Meteoritical Bulletin Database (https://www.lpi.usra.edu/meteor/, accessed March 2024, N = 1749). Data from ungrouped achondrites (blue points in panel b and c) suggest (i) distinct parent bodies in the fields typical of CC and NC achondrites, and (ii) underexplored links between meteorites sampling different reservoirs within the same parent body.

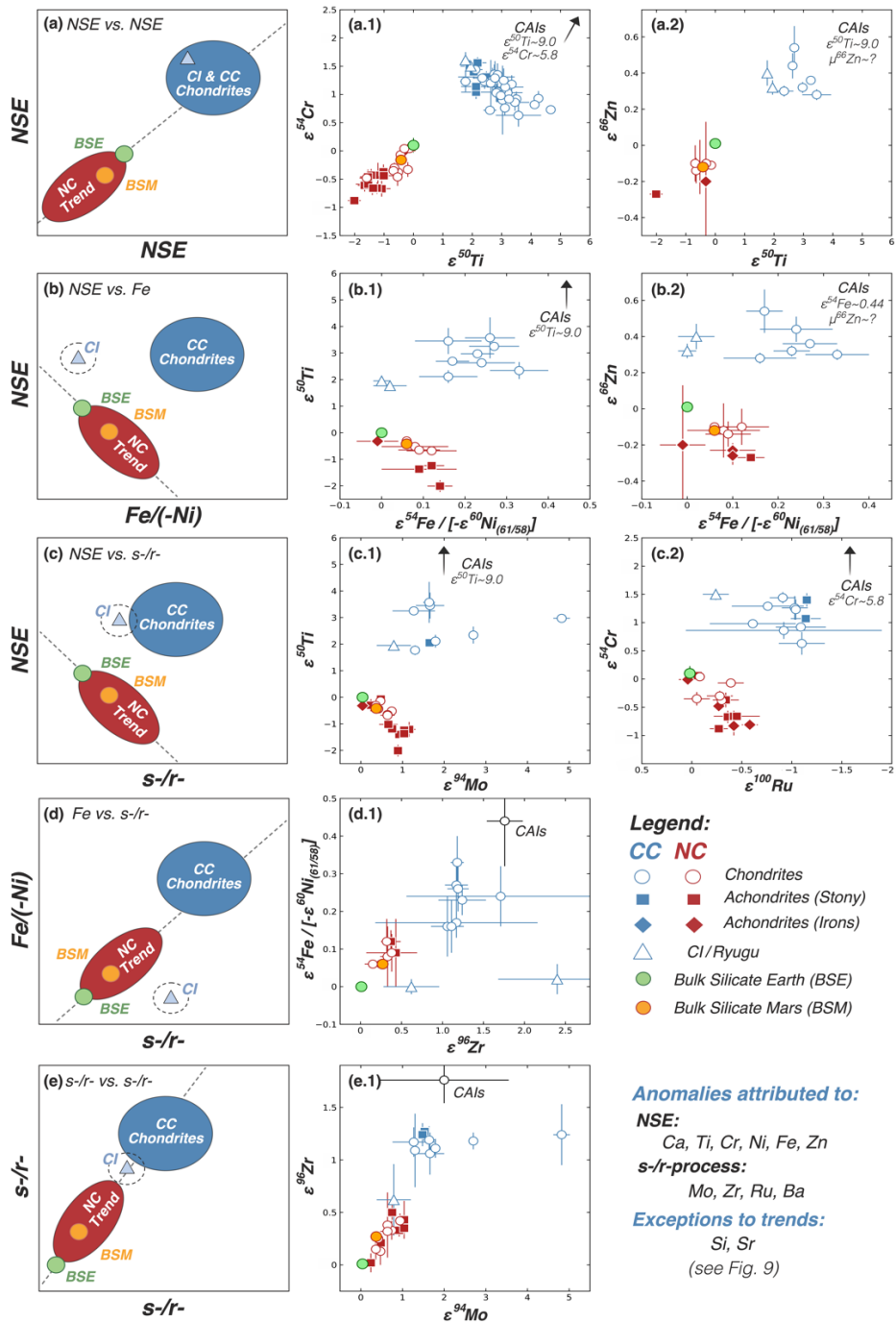

**Fig. 8 Isotope anomaly systematics of NC and CC meteorites and silicate portions of Earth (BSE) and Mars (BSM) in multielement space.** NC meteorites representing the inner SS are shown in red, and CC meteorites representing the outer SS are shown in blue. Note that in plots with $\varepsilon^{100}$Ru, we invert the axis such that enrichment in supernova-derived nuclides (towards CAIs) is directed to the top right, as in all other plots. Owing to the conventionally adopted internal normalization scheme for Ru anomalies, higher $\varepsilon^{100}$Ru reflects greater *s*-process contributions. See Supplementary Materials for data source.

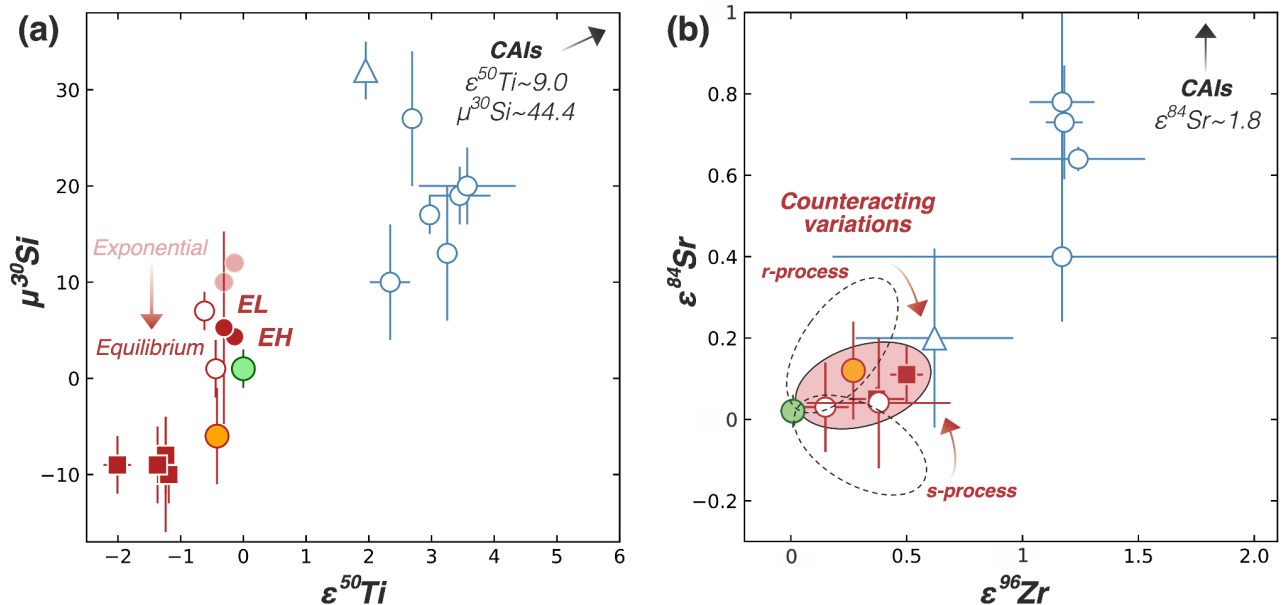

**Fig. 9. Isotope anomalies of (a) Si vs Ti, and (b) Sr vs Zr.** At first glance, both Si and Sr do not seem to conform to the general trends observed for other elements (see Fig. 8). (a) The elevated Si isotope anomalies observed in enstatite chondrites may be due to improper use of the exponential law to correct for MDF, which is expected to follow the equilibrium law at the high-T applicable to the early solar nebula. (b) The undistinguishable Sr isotope anomalies in NC materials, which contrast with the clear trends in NC materials observed for other elements, is thought to testify to correlated *s*- and *r*-/*p*-anomalies, whose effect cancel out. See main text for further information. Symbols as in Fig. 8.

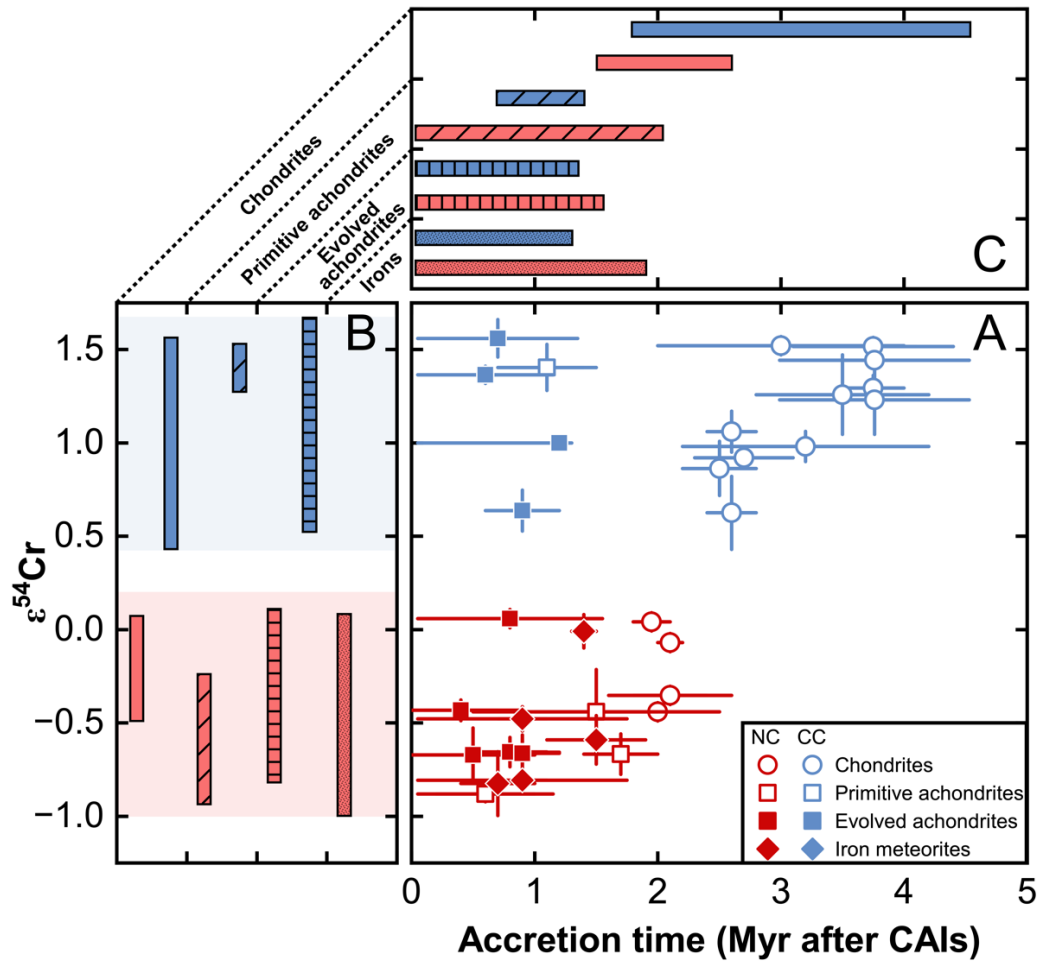

**Fig. 10. Diagram of chromium isotopic composition versus inferred accretion ages of various meteorite groups (a) and bar diagrams of $\varepsilon^{54}$Cr (b) and inferred accretion ages (c) of NC and CC meteorites individually.** Panel c) indicates that iron meteorite and stony achondrite parent body accretion started rapidly and coevally within the first ~2 Myr after CAIs in both reservoirs followed by the formation of chondrites at ~2–4 Myr after CAIs. Importantly, the isotopic dichotomy was already established within ~1 Myr after CAIs (a). Furthermore, iron meteorites, stony achondrites, and chondrites appear to cover almost the entire isotopic range of both reservoirs (b) arguing against a temporal evolution for the overall isotopic evolution of the NC and CC reservoirs. Red and blue shaded areas in b) indicate the full isotopic range of NC and CC meteorites, respectively. The size of the bars in b) and c) include the respective uncertainties on the sample values of a certain group. Data as reported in Tables S2-3.

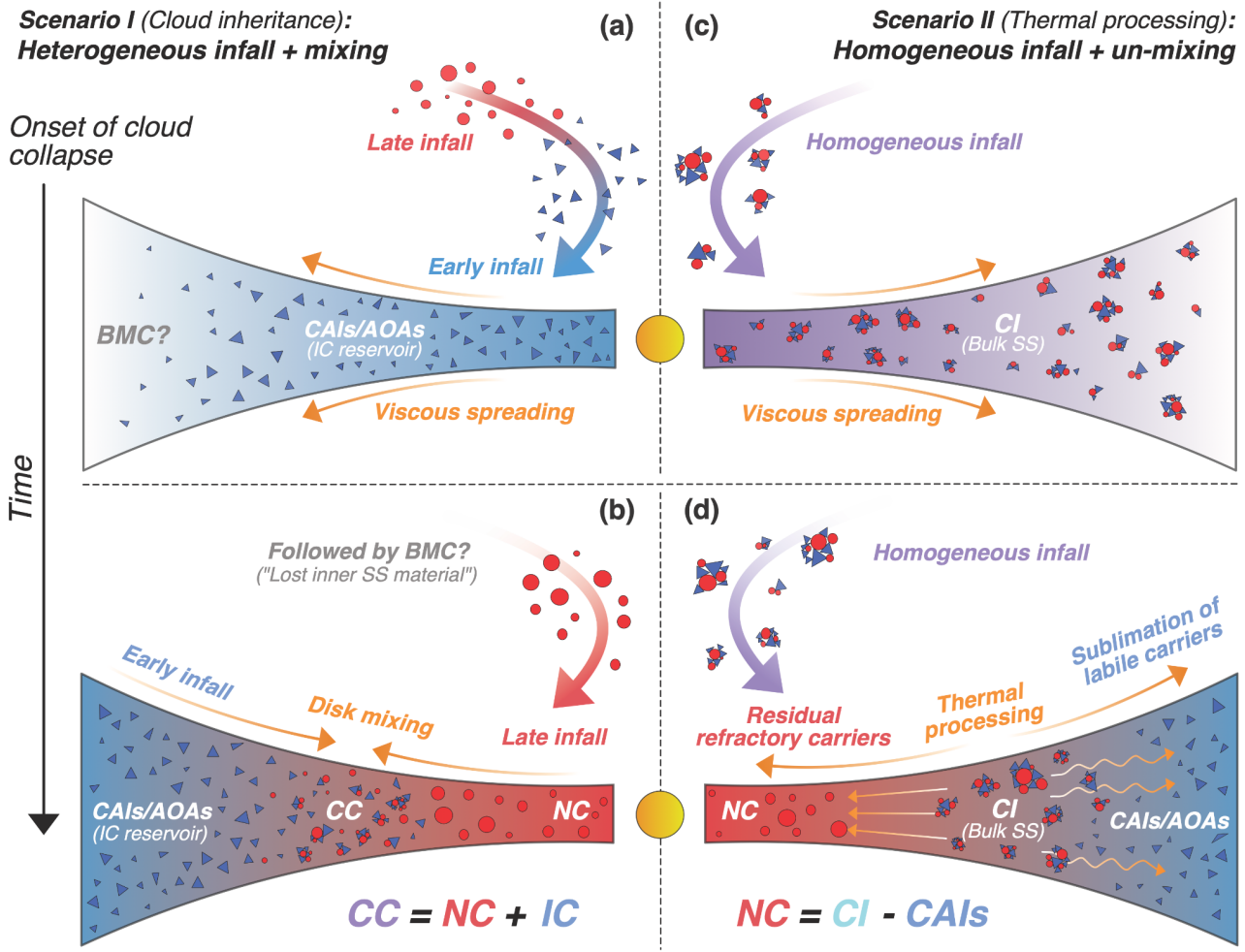

**Fig. 11. Schematic representation of the two families of models that have been proposed to explain the isotopic heterogeneity observed in SS materials.** In (a,b) the NC-CC dichotomy reflects the inheritance of isotopic heterogeneity initially present in the molecular cloud, and CC bodies constitute mixtures of isotopically NC and Inclusion-like Chondritic (IC) materials, where IC denotes an hypothetical reservoir that is isotopically CAI-like but chemically chondritic (*e.g.*, Nanne et al, 2019; Burkhardt et al., 2019; Yap & Tissot et al., 2023). In (c,d) the presolar carriers were homogeneously distributed within the protosolar nebula (CI-like bulk SS), and isotopic heterogeneity arose from sorting of presolar carriers through thermal and physical processing within the disk (*e.g.*, Trinquier et al., 2009; Dauphas et al., 2010; Burkhardt et al., 2012; Davis et al., 2018).

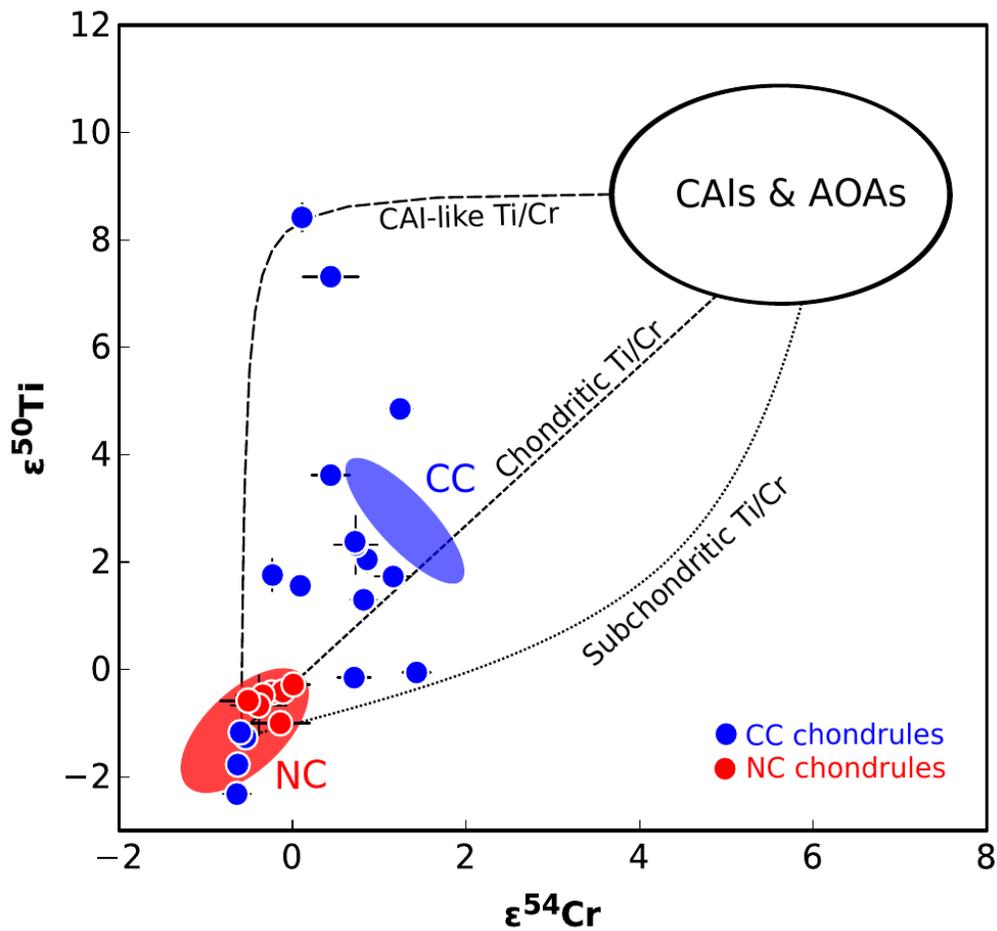

**Fig. 12. ε⁵⁴Cr versus ε⁵⁰Ti isotopic anomalies of individual NC and CC chondrules and their relation to the composition of bulk NC and CC bodies and refractory inclusion.** While the isotopic composition of NC chondrules is in line with the composition of bulk NC bodies the CC chondrules span a range of compositions from NC to CAI-like. This might be explained by variable mixing of CAIs and AOAs in the CC chondrule precursors. Curves represent mixing lines between NC materials and material with an isotopic composition of CAIs and AOAs. The curvature of the mixing lines depends on the Ti/Cr element ratio of the mixing endmember. Data from Schneider et al., (2020) and Williams et al., (2020).

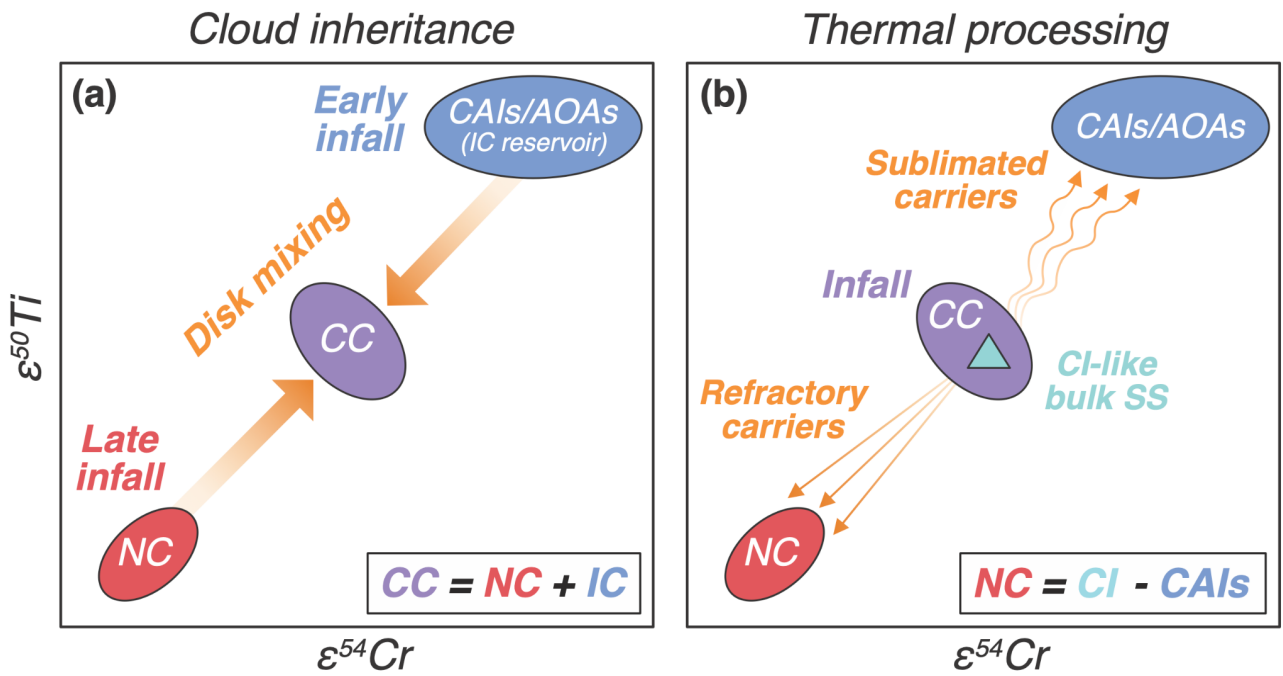

**Fig. 13. Schematic representation in isotope-isotope space of the two families of models that have been proposed to explain the isotopic heterogeneity observed in SS materials.** For a description of each model, see Fig. 11.

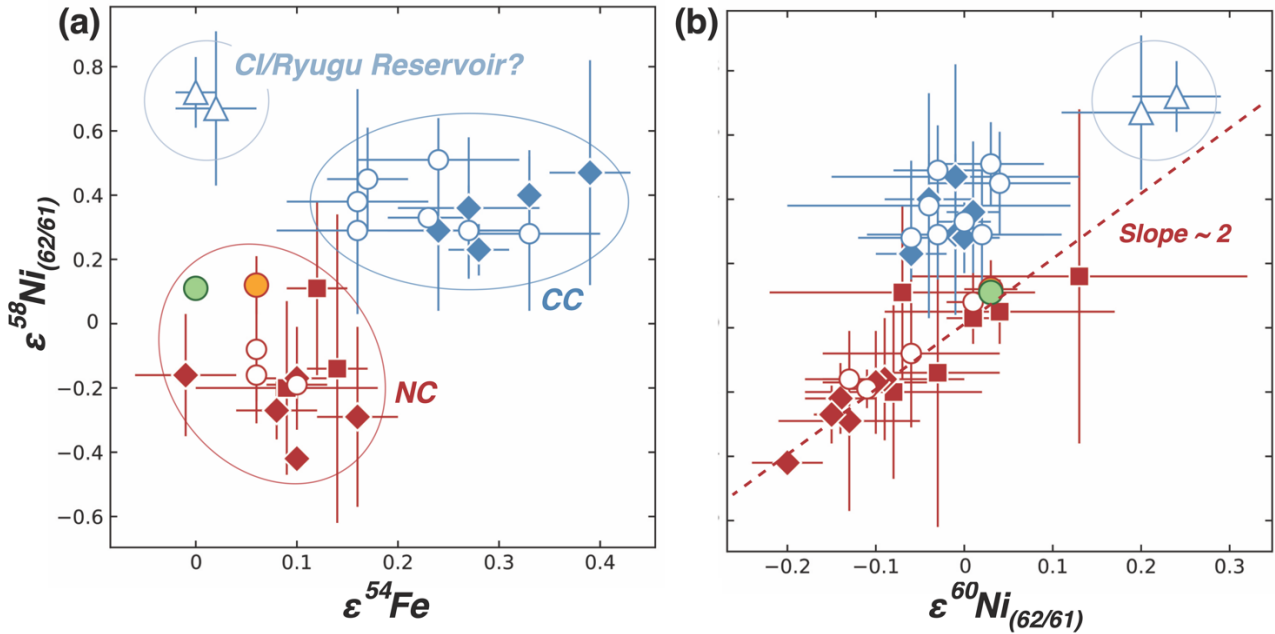

**Fig. 10. Isotope anomalies of (a) $^{58}$Ni vs $^{54}$Fe, and (b) $^{58}$Ni vs $^{60}$Ni.** Nickel anomalies are internally normalized to the terrestrial $^{62}$Ni/$^{61}$Ni ratio (see subscript). At current limits of precision, CI chondrites and Ryugu (blue triangles) exhibit identical Fe and Ni anomalies, distinct from those of other CC chondrites. In particular, CIs/Ryugu are characterized by a depletion in $^{54}$Fe, and excess in $^{58}$Ni and $^{60}$Ni, relative to the other CCs. This observation suggests the parent bodies of CIs and Ryugu incorporated a unique carrier (*e.g.,* micron-sized Fe-Ni grains; Spitzer et al. 2024) during their formation, and underlies the (presently debated) proposition that they derive from a third isotopic reservoir in the early SS (Hopp et al. 2022a; *see Section 4.3*). Symbols as in Fig. 8.

**Table 1. Cosmochemical classification of the elements for which nucleosynthetic isotope anomalies have been identified in bulk meteorites.**

| Elements | Lithophile (silicate + oxides) | Siderophile/ chalcophile (metal + sulfides) |
|---|---|---|
| Refractory ($T_c$ 1850–1355 K) | Zr, Nd, Sm, **Ti**, **Ca**, Sr, Ba | Os, W, Mo, Ru, Pt |
| Moderately refractory ($T_c$ 1355–1250 K) | **Mg**, **Si**, Cr | **Ni**, **Fe**, Pd |
| Moderately volatile ($T_c$ 1250–252 K) | K, Zn | |
| Highly volatile ($T_c$ < 252 K) | (O), Ne, Xe | |

*Elements in order of decreasing condensation temperatures ($T_C$) at a pressure of $10^{-4}$ bar (Lodders, 2003). Major elements are shown in boldface. Modified after (McDonough, 2016).*